\documentclass[seceq,preprint]{ptptex}

\usepackage{graphicx}

\newcommand{\wt}{\widetilde}
\newcommand{\wh}{\widehat}
\newcommand{\ol}{\overline}
\newcommand{\del}{\partial}
\newcommand{\ra}{\rightarrow}

\newcommand{\lra}{\leftrightarrow}
\newcommand{\nn}{\nonumber}
\newcommand{\half}{\frac{1}{2}}

\def\tr{\mathop{\rm tr}\nolimits}

\newcommand{\AD}[1]{$\ol{\mbox{D~\,}}\!\!\!$#1}
\newcommand{\gym}{g_{\rm Y\!M}}
\newcommand{\cA}{{\cal A}}
\newcommand{\cL}{{\cal L}}
\newcommand{\cV}{{\cal V}}
\newcommand{\bp}{\mbox{\boldmath $p$}}
\newcommand{\bq}{\mbox{\boldmath $q$}}
\newcommand{\bR}{\mathbb{R}}

\newcommand{\cO}{{\cal O}}

\newcommand{\Ukk}{U_{\rm KK}}
\newcommand{\Mkk}{M_{\rm KK}}

\notypesetlogo                       

\title{
More on a Holographic Dual of QCD%
}

\author{
Tadakatsu \textsc{Sakai}$^{1,}$\footnote{E-mail:
tsakai@mx.ibaraki.ac.jp}
and
Shigeki \textsc{Sugimoto}$^{2,}$\footnote{E-mail:
sugimoto@yukawa.kyoto-u.ac.jp}
}

\inst{
$^1$Department of Mathematical Sciences, Faculty of Sciences,\\
Ibaraki University, Mito 310-8512, Japan\\ 
$^2$Yukawa Institute for Theoretical Physics, Kyoto
University,\\ Kyoto 606-8502, Japan
}

\abst{
We investigate the interactions among the pion, vector mesons and
external gauge fields in the holographic dual of massless QCD
proposed in a previous paper, Ref.~\citen{SaSu}, on the basis of
probe D8-branes embedded in a D4-brane
background in type IIA string theory.
We obtain the coupling constants by performing
both analytic and numerical calculations,
and compare them with experimental data.
It is found that the vector meson dominance in the pion
form factor as well as in the Wess-Zumino-Witten term holds
in an intriguing manner. 
We also study the $\omega\ra\pi\gamma$
and $\omega\ra 3\pi$ decay amplitudes.
It is shown that the interactions relevant to these
decay amplitudes have the same
structure as that proposed 
by Fujiwara et al.\cite{FuKuTeUeYa}
Various relations among the masses and the coupling constants
of an infinite tower of mesons are derived. These relations
play crucial roles in the analysis.
We find that most of the results are consistent with
experiments.
}

\begin{document}

\maketitle

\section{Introduction}

In a previous paper, Ref.~\citen{SaSu}, we proposed a holographic dual of 
$U(N_c)$ QCD with $N_f$ massless flavors,
which is constructed by putting
probe D8-branes in the D4-brane background.
It was shown there that various phenomena that are expected to occur
in low energy QCD can be reproduced in this framework.
For instance, we showed that the chiral $U(N_f)_L\times U(N_f)_R$
symmetry is spontaneously broken to the diagonal subgroup $U(N_f)_V$.
The associated Nambu-Goldstone (NG) bosons were found
and identified with the pion. Moreover, we 
found vector mesons in the spectrum, and the masses and some of the
coupling constants among them turn out to be reasonably close to the
experimental values.

The purpose of this paper is to study the D4/D8 model in more detail
in order to explore the low-energy phenomena involving the mesons.
The effective action of our model consists of two parts. One
is the five-dimensional Yang-Mills (YM) action on a curved background, 
which originates from the non-abelian Dirac-Born-Infeld (DBI)
action on the probe. The other is the integral of the Chern-Simons (CS)
five-form, which results from the CS term on
the probe D8-brane.
{}From these, we compute the cubic and some quartic interaction terms
among the pion, the vector mesons and the external gauge fields
associated with the chiral $U(N_f)_L\times U(N_f)_R$ symmetry.
The results are compared with the experimental data
in order to quantitatively test the conjectured duality.
For recent developments toward holographic descriptions of QCD, see
also Refs.~\citen{SoSt}-\citen{Hirn:2005nr}.

In particular, we are interested in the coupling to the external
photon field. We examine whether the vector meson dominance
hypothesis \cite{GeZa,Sakurai} is satisfied in this model.
This hypothesis states that
the exchange of vector mesons dominates the
electromagnetic interactions of hadrons.
For example, the electromagnetic form factor of the pion
is dominated by the $\rho$ meson pole as
\begin{eqnarray}
F_\pi(k^2)\simeq\frac{g_{\rho}g_{\rho\pi\pi}}{k^2+m_{\rho}^2}\ ,
\end{eqnarray} 
where $g_\rho$ is the $\rho$ meson decay constant, 
$m_\rho$ is the $\rho$ meson mass and $g_{\rho\pi\pi}$
is the $\rho\,\pi\pi$ coupling.
In other words, the direct couplings between the photon and
the pion are small compared with
the indirect interactions resulting from the $\rho$ meson
exchange. It has been shown in
Refs.~\citen{SoSt,HoYoSt1,HoYoSt2} and \citen{DaRoPo}
that the pion form factor exhibits
vector meson dominance in generic holographic models of QCD,
where the contributions from infinitely many vector mesons
are important.
We reexamine this feature in our model and present a numerical
estimation of the dominant terms. Furthermore,
we analyze the Wess-Zumino-Witten (WZW) term
that includes an infinite tower of vector mesons
and demonstrate the complete vector meson dominance in this sector.

The subjects considered in this paper also include
the Kawarabayashi-Suzuki-Riazuddin-Fayyazuddin
(KSRF) relations \cite{KS,RF}, the pion charge radius,
$a_1\ra\pi\,\gamma$ and $a_1\ra\pi\rho$ decay,
$\pi\pi$ scattering, the Weinberg sum rules \cite{W:sum}, and
$\omega\ra\pi^0\gamma$ and $\omega\ra \pi^0\pi^+\pi^-$ decay.
{}For most of the cases, 
we obtain considerably good agreement with the experimental data.

This paper is organized as follows.
In \S \ref{rev}, we review the D4/D8 model to the extent
needed in this paper. Note that the notation used in
this paper is slightly different from that used in Ref.~\citen{SaSu}.
We define our notation in this section.
In \S \ref{DBIpart}, we investigate the DBI part of the model.
Section \ref{WZWterm} is devoted to analyzing the WZW term.
In \S \ref{revisit}, we reanalyze the effective action using a different
gauge, which simplifies the treatment of the vector meson dominance.
We end this paper with summary and
discussion in \S \ref{discussion}.
The two appendices summarize some technical computations.

\section{The model}
\label{rev}
In this section,
we review the D4/D8 model proposed in Ref.~\citen{SaSu}
and define the notation used in this paper.

The D4/D8 model is formulated by placing probe D8-branes
into the D4-brane background proposed in Ref.~\citen{Witten;thermal}
as a supergravity dual of four-dimensional
$U(N_c)$ Yang-Mills theory.
The metric, dilaton $\phi$, and the RR three-form field
$C_3$ in the D4-brane background are given as
\begin{eqnarray}
&&ds^2=\left(\frac{U}{R}\right)^{3/2}
\left(\eta_{\mu\nu}dx^\mu dx^\nu+f(U)d\tau^2\right)
+\left(\frac{R}{U}\right)^{3/2}
\left(\frac{dU^2}{f(U)}+U^2 d\Omega_4^2\right),
\nn\\
&&~~~~e^\phi= g_s \left(\frac{U}{R}\right)^{3/4},
~~F_4\equiv dC_3=\frac{2\pi N_c}{V_4}\epsilon_4 \ ,
~~~f(U)\equiv 1-\frac{\Ukk^3}{U^3} \ .
\label{D4sol}
\end{eqnarray}
Here the coordinates $x^\mu$ ($\mu=0,1,2,3$)
and $\tau$ parameterize the directions
along which the D4-brane is extended, and
$U$ corresponds to the radial direction transverse
to the D4-brane.
{}From the definition of the function $f(U)$, we see that $U$ is
bounded from below as $U\ge \Ukk$. The quantities
$d\Omega_4^2$, $\epsilon_4$ and $V_4=8\pi^2/3$
are the line element,
the volume form, and the volume of a unit $S^4$ surrounding
the D4-brane, respectively, and
 $R$ and $\Ukk$ are constant parameters. The constant $R$ is related
to the string coupling $g_s$ and the string length $l_s$ as
$R^3=\pi g_s N_c l_s^3$. This background represents
$N_c$ D4-branes wrapped on a supersymmetry breaking $S^1$
parameterized by the parameter $\tau$, whose period is chosen as
\begin{eqnarray}
\tau\sim\tau+2\pi \Mkk^{-1}\ ,~~~~
\Mkk\equiv\frac{3}{2}\frac{\Ukk^{1/2}}{R^{3/2}} \ ,
\end{eqnarray}
in order to avoid a conical singularity at $U=\Ukk$.
Along this $S^1$, fermions are taken to be anti-periodic,
and they become massive as four-dimensional fields.
Adjoint scalar fields on the D4-brane
are also expected to acquire a mass via
quantum effects, since the supersymmetry is completely broken.
Thus, the world-volume theory on the D4-brane effectively
becomes the four-dimensional Yang-Mills theory
below the Kaluza-Klein mass scale $\Mkk$.
The Yang-Mills coupling $\gym$ (at the scale $\Mkk$)
is given by $\gym^{2}=2\pi\Mkk g_s l_s$, which
is read off of the DBI action of the D4-brane compactified on $S^1$.
The parameters $R$, $\Ukk$ and $g_s$ are expressed in terms
of $\Mkk$, $\gym$ and $l_s$. 
One can easily show that $l_s$ does not appear
in the effective action if it is written in terms of $\Mkk$ and $\gym$.
Therefore, without loss of generality, we can set
\begin{eqnarray}
\frac{2}{9}\,\Mkk^2 l_s^2=(\gym^2 N_c)^{-1}\equiv \lambda^{-1}\ ,
\label{alpha}
\end{eqnarray}
which makes $R$ and $\Ukk$ independent of $\gym$ and $N_c$.
Furthermore, because the $\Mkk$ dependence is easily recovered
through dimensional analysis, it is convenient to work in units
in which $\Mkk=1$.
Then, we have the relations
\begin{eqnarray}
\Mkk=1\ ,~~~
R^3=\frac{9}{4}\ ,~~~\Ukk=1\ ,~~~
\frac{1}{g_sl_s^3}=\frac{4\pi}{9} N_c\ .
\label{RUg}
\end{eqnarray}
The relations (\ref{alpha}) and (\ref{RUg}) make it clear that
the $\alpha'$ expansion and the loop expansion in string theory
correspond to the expansion
with respect to $1/\lambda$ and $\lambda^{3/2}/N_c$
in Yang-Mills theory, respectively.
In this paper, we consider only the leading terms in this
expansion by taking $N_c$ and $\lambda$ to be
sufficiently large.

In order to add $N_f$ flavors of quarks to the supergravity dual
of the Yang-Mills theory described by the background (\ref{D4sol}),
we place $N_f$ probe D8-branes extended along 
 $x^\mu$ ($\mu=0,1,2,3$), the $S^4$ directions, and
one of the directions in the $(U,\tau)$ plane.
Here we adopt the probe approximation,
assuming $N_c\gg N_f$, and
ignore the backreaction from the D8-branes to the D4-brane
background.
To describe the D8-branes, it is convenient to introduce
new coordinates $(y,z)$ defined by
\begin{eqnarray}
(y,z)\equiv\left(\sqrt{U^3-1}\cos\tau\,, \sqrt{U^3-1}\sin\tau\right)\ .
\label{yz}
\end{eqnarray}
It is easy to show that the metric written in $(y,z)$ is
smooth everywhere.
We consider the probe D8-branes placed at $y=0$ and extended
along the $z$ direction.
As discussed in Ref.~\citen{SaSu},
this brane configuration corresponds to
a D4/D8/\AD8 system that represents $U(N_c)$ QCD
with $N_f$ massless flavors.

Note that this system possesses $SO(5)$ symmetry
corresponding to the rotations of $S^4$.
In this paper, we concentrate on
the states that are invariant under $SO(5)$ rotations
for simplicity. Because QCD does not have such an $SO(5)$ symmetry,
the meson in realistic QCD can only be found in this
sector.\footnote{
It is believed somewhat optimistically that
states charged under $SO(5)$ decouple in the $\Mkk\ra\infty$
limit. (See Ref.~\citen{adscft} and references therein.)
See also Ref.~\citen{PoRuTa} for a recent analysis.}
Therefore, we can reduce the nine-dimensional gauge theory on
the D8-brane to a five-dimensional theory with
a five-dimensional $U(N_f)$ gauge field denoted by
$A_\mu(x^\mu,z)$ and $A_z(x^\mu,z)$.\footnote{Here we omit
the scalar field $y(x^\mu,z)$, which corresponds
to fluctuations of the D8-brane along the transverse direction,
for simplicity.}

The effective action on the probe D8-brane embedded 
in the background (\ref{D4sol}) consists of two parts.
One is the (non-Abelian) DBI action, and the other is
the CS term.
After the Kaluza-Klein reduction on $S^4$,
the leading terms in the $1/\lambda$ expansion
of the DBI action read
\begin{eqnarray}
S_{\rm D8}^{\rm DBI}=
\kappa\int d^4 x dz \,
\tr\left[\,
\half K^{-1/3} F_{\mu\nu}^2+ KF_{\mu z}^2
\,\right]\ ,
\label{DBI}
\end{eqnarray}
where
\begin{eqnarray}
\kappa\equiv\frac{\lambda N_c}{216\pi^3}\ ,~~~K(z)\equiv 1+z^2\ .
\end{eqnarray}
The CS term is 
\begin{eqnarray}
S_{\rm D8}^{\rm CS}=\frac{N_c}{24\pi^2}\int_{M^4\times\bR}
\omega_5(A)\ ,
\label{CS}
\end{eqnarray}
where $\omega_5(A)$ is the Chern-Simons five-form written in terms of
the five-dimensional differential form $A=A_\mu d x^\mu+A_z d z$ as
\begin{eqnarray}
\omega_5(A)=\tr\left(
AF^2-\half A^3 F+\frac{1}{10}A^5
\right) \ ,
\end{eqnarray}
and $M^4\times\bR$ is the five-dimensional space-time parameterized
by $(x^\mu,z)$.

In order to extract four-dimensional meson fields from the
five-dimensional gauge field, we expand the gauge field as
\begin{eqnarray}
A_\mu(x^\mu,z)&=&
\sum_{n=1}^\infty B_\mu^{(n)}(x^\mu)\psi_n(z)\ ,
\label{exp1}\\
A_z(x^\mu,z)&=&\varphi^{(0)}(x^\mu)\phi_0(z)+
\sum_{n=1}^\infty \varphi^{(n)}(x^\mu)\phi_n(z)\ ,
\label{exp2}
\end{eqnarray}
using the complete sets $\{\psi_n(z)\}_{n\ge 1}$
and $\{\phi_n(z)\}_{n\ge 0}$ of functions of $z$.
In order to diagonalize
the kinetic terms and the mass terms of
the four-dimensional fields $B^{(n)}_\mu(x^\mu)$ and
$\varphi^{(n)}(x^\mu)$, we choose the functions
$\psi_n(z)$ to be eigenfunctions satisfying the equation
\begin{equation}
-K^{1/3}\,\del_z\left(
K\,\del_z\psi_n\right)=
\lambda_n\psi_n \ ,
\label{deqn;psi}
\end{equation}
where $\lambda_n$ is the eigenvalue, and
the normalization condition is taken to be
\begin{eqnarray}
\kappa\int dz \,K^{-1/3}\psi_n\psi_m&=&\delta_{nm}\ .
\label{norm;psi}
\end{eqnarray}
The functions $\phi_n(z)$ are chosen to satisfy
 $\phi_n(z)\propto \del_z\psi_n(z)$ ($n\ge 1$)
 and $\phi_0(z)=1/(\sqrt{\pi\kappa}K(z))$,
with the normalization condition
\begin{eqnarray}
\kappa\int dz \,K\phi_n\phi_m=\delta_{nm}\ ,
\end{eqnarray}
which is compatible with (\ref{deqn;psi}) and (\ref{norm;psi}).

Inserting the expansion consisting of
(\ref{exp1}) and (\ref{exp2}) into the action
(\ref{DBI}) and integrating over $z$, we obtain
\begin{eqnarray}
S^{\rm DBI}_{\rm D8}&\simeq&
\int d^4 x\,\tr\Bigg[\,
(\del_\mu\varphi^{(0)})^2\nn\\
&&~~+\sum_{n=1}^\infty\left(
\half
(\del_\mu B^{(n)}_\nu-\del_\nu B^{(n)}_\mu)^2
+\lambda_n
(B_\mu^{(n)}-\lambda_n^{-1/2}\del_\mu\varphi^{(n)})^2
\right)\Bigg]\nn\\
&&~~+(\mbox{interaction terms})\ .
\end{eqnarray}
{}From this, we see that we have one massless scalar field,
$\varphi^{(0)}$, and a tower of massive vector fields, $B_\mu^{(n)}$,
of mass squared $\lambda_n$. The scalar fields $\varphi^{(n)}$
with $n\ge 1$ are eaten by the vector fields $B^{(n)}_\mu$.
We interpret $\varphi^{(0)}$ as the massless pion field and
 $B^{(n)}_\mu$ as vector meson fields.

In the expansion given in (\ref{exp1}) and (\ref{exp2}), we have
implicitly assumed that the gauge field asymptotically vanishes 
$A_M(x^\mu,z)\ra 0$
as $z\ra\pm\infty$. The residual gauge transformation 
that does not violate this condition is obtained with
a gauge function $g(x^\mu,z)$ that asymptotically
becomes constant: $g(x^\mu,z)\ra g_\pm$ as $z\ra \pm\infty$.
We interpret $(g_+,g_-)$ as an element of the chiral symmetry
group $U(N_f)_L\times U(N_f)_R$ in QCD with $N_f$ massless flavors.

In the following sections, we study the interaction of the
mesons with the external gauge fields $(A_{L\mu},A_{R\mu})$ introduced
by weakly gauging the $U(N_f)_L\times U(N_f)_R$ chiral symmetry.
Of particular interest are 
the couplings of the mesons to the photon field $A^{\rm em}_\mu$,
which can be extracted by setting
\begin{eqnarray}
A_{L\mu}=A_{R\mu}= e Q A_\mu^{\rm em}\ ,
\end{eqnarray}
where $e$ is the electromagnetic coupling constant and
$Q$ is the electric charge matrix given, for example, by
\begin{eqnarray}
Q=\frac{1}{3}\left(
\begin{array}{ccc}
2\\
&-1\\
&&-1
\end{array}
\right)\ ,
\end{eqnarray}
for the $N_f=3$ case.
It is also necessary to introduce the external gauge fields
in the calculation of correlation functions
among the currents associated with the chiral
symmetry following the prescription
used in the AdS/CFT correspondence \cite{GKP,Witten:ads}.
In order to turn on the external gauge fields, we impose
the asymptotic values of the gauge field $A_\mu$ on the
D8-brane as
\begin{eqnarray}
\lim_{z\ra+\infty}
A_\mu(x^\mu,z)= A_{L\mu}(x^\mu)\ ,~~~
\lim_{z\ra-\infty}
A_\mu(x^\mu,z)= A_{R\mu}(x^\mu)\ .
\label{asymp}
\end{eqnarray}
This is implemented by modifying the mode expansion (\ref{exp1}) as
\begin{eqnarray}
A_\mu(x^\mu,z)&=&
A_{L\mu}(x^\mu)\psi_+(z)+A_{R\mu}(x^\mu)\psi_-(z)+
\sum_{n=1}^\infty B_\mu^{(n)}(x^\mu)\psi_n(z)\ ,
\label{extexp}
\end{eqnarray}
where the functions $\psi_\pm(z)$ are defined as
\begin{eqnarray}
\psi_\pm(z)\equiv\half(1\pm\psi_0(z))\ ,~~~
\psi_0(z)\equiv \frac{2}{\pi}\arctan z \ ,
\end{eqnarray}
which 
are the non-normalizable zero modes of (\ref{deqn;psi})
satisfying $\del_z\psi_\pm(z)\propto\phi_0(z)$.

Note that
if we insert the expansion (\ref{extexp}) into
the action (\ref{DBI}) and perform the integration over $z$,
the coefficients of the kinetic terms of the gauge fields
$A_{L\mu}$ and $A_{R\mu}$ diverge,
because $\psi_\pm$ are non-normalizable.
This divergence simply reflects the fact that
the gauge coupling corresponding
to the chiral $U(N_f)_L\times U(N_f)_R$ symmetry is zero.
One way to regularize the divergence
is to cut off the integration over $z$ at some large but finite value.
Another possibility is to simply ignore the divergent
kinetic terms of the external gauge field, since we are
interested only in the structure of the interactions.

In this paper, we work mainly in the $A_z=0$ gauge,
which can be realized by applying the gauge transformation
$A_M\ra g A_M g^{-1}+ g\del_M g^{-1}$
with the gauge function
\begin{eqnarray}
g^{-1}(x^\mu,z)= P \exp\left\{-
\int_{0}^z dz'\, A_z(x^\mu,z')
\right\} \ .
\label{ginv}
\end{eqnarray}
Then, the asymptotic values (\ref{asymp}) change to
\begin{eqnarray}
\lim_{z\ra+\infty}
A_\mu(x^\mu,z)= A_{L\mu}^{\xi_+}(x^\mu)\ ,~~~
\lim_{z\ra-\infty}
A_\mu(x^\mu,z)= A_{R\mu}^{\xi_-}(x^\mu)\ ,
\end{eqnarray}
where
$\xi_\pm(x^\mu)\equiv \lim_{z\ra\pm\infty}g(x^\mu,z)$
and
\begin{eqnarray}
A_{L\mu}^{\xi_+}(x^\mu)&\equiv&
\xi_+(x^\mu)A_{L\mu}(x^\mu)\xi_+^{-1}(x^\mu)+
\xi_+(x^\mu)\del_\mu\xi_+^{-1}(x^\mu)\ ,\\
A_{R\mu}^{\xi_-}(x^\mu)&\equiv&
\xi_-(x^\mu)A_{R\mu}(x^\mu)\xi_-^{-1}(x^\mu)+
\xi_-(x^\mu)\del_\mu\xi_-^{-1}(x^\mu)\ .
\end{eqnarray}
Then, the gauge field in the $A_z=0$ gauge can be expanded as
\begin{eqnarray}
A_\mu(x^\mu,z)&=&
A_{L\mu}^{\xi_+}(x^\mu)\psi_+(z)
+A_{R\mu}^{\xi_-}(x^\mu)\psi_-(z)
+\sum_{n=1}^\infty B_\mu^{(n)}(x^\mu)\psi_n(z)\ .
\label{5dpot}
\end{eqnarray}

The residual gauge symmetry in the $A_z=0$ gauge
is given by the $z$-independent gauge transformation.
The residual gauge symmetry $h(x^\mu)\in U(N_f)$ and
the weakly gauged chiral symmetry
$(g_+(x^\mu),g_-(x^\mu))\in U(N_f)_L\times U(N_f)_R$
act on these fields as
\begin{eqnarray}
A_{L\mu}&\ra& g_+A_{L\mu}g_+^{-1}+g_+\del_\mu g_+^{-1}\ ,
\label{tr1}\\
A_{R\mu}&\ra& g_-A_{R\mu}g_-^{-1}+g_-\del_\mu g_-^{-1}\ ,\\
\xi_\pm&\ra& h\,\xi_\pm\, g_\pm^{-1}\ ,\label{xih}\\
B_\mu^{(n)}&\ra& h\,B_\mu^{(n)}\, h^{-1}\ .
\label{tr2}
\end{eqnarray}
Here, the functions $\xi_\pm(x^\mu)$ are interpreted as 
the $U(N_f)$ valued fields $\xi_{L,R}(x^\mu)$
that carry the pion degrees of freedom in the 
hidden local symmetry approach \cite{BKUYY,bky;review}.
Actually, the transformation property (\ref{xih}) is
the same as that for $\xi_{L,R}(x^\mu)$
if we interpret $h(x^\mu)\in U(N_f)$ as the hidden local symmetry.
These fields are related to the $U(N_f)$ valued
pion field $U(x^\mu)$ in the chiral Lagrangian by
\begin{eqnarray}
\xi_+^{-1}(x^\mu)\xi_-(x^\mu)=U(x^\mu)\equiv
e^{2i\Pi(x^\mu)/f_\pi} \ .
\end{eqnarray}
The pion field $\Pi(x^\mu)$ is identical to $\varphi^{(0)}(x^\mu)$
in (\ref{exp2}) up to linear order.
\footnote{
Here we take the pion field $\Pi(x^\mu)$ to be a Hermitian matrix,
while $\varphi^{(0)}(x^\mu)$ and
the vector meson fields $B_\mu^{(n)}(x^\mu)$ are anti-Hermitian.
}

Choosing $h(x^\mu)$ in (\ref{xih}) appropriately,
we can choose the gauge such that
\begin{eqnarray}
\xi_+^{-1}(x^\mu)=\xi_-(x^\mu)=e^{i\Pi(x^\mu)/f_\pi}\ .
\end{eqnarray}
In this gauge, the gauge potential in (\ref{5dpot}) can be
expanded up to quadratic order in the fields as
\begin{eqnarray}
A_\mu
&=&
\left(\cV_\mu+\frac{1}{2f_\pi^2}[\,\Pi,\del_\mu\Pi\,]
-\frac{i}{f_\pi}[\,\Pi,\cA_\mu\,]\right)
+\left(\cA_\mu+\frac{i}{f_\pi}\del_\mu\Pi
-\frac{i}{f_\pi}[\,\Pi,\cV_\mu\,]\right)\psi_0
\nn\\
&&~~~+\sum_{n=1}^\infty v^n_\mu\,\psi_{2n-1}
+\sum_{n=1}^\infty a^n_\mu\,\psi_{2n}+\cdots\ ,
\label{exp}
\end{eqnarray}
with
\begin{eqnarray}
\cV_\mu\equiv\half(A_{L\mu}+A_{R\mu})\ ,~
\cA_\mu\equiv\half(A_{L\mu}-A_{R\mu})\ ,~
v^n_\mu\equiv B_\mu^{(2n-1)}\ ,~
a^n_\mu\equiv B_\mu^{(2n)}\ .
\label{VAva}
\end{eqnarray}
Note that the functions $\psi_n(z)$ are even and odd functions of
$z$ for odd and even values of $n$, respectively. This implies that
$v^n$ and $a^n$ are vector and axial-vector mesons, respectively.
As discussed in Ref.~\citen{SaSu}, the lightest vector meson, $v^1$, is
interpreted as the $\rho$ meson [$\rho(770)$]
and the lightest axial vector meson,
$a^1$, is interpreted as the $a_1$ meson [$a_1(1260)$].
The fields $v^2,v^3,\cdots$ and
$a^2,a^3,\cdots$ represent the heavier vector and axial-vector
mesons with the same quantum numbers:
$\rho(1450),\rho(1700),\cdots$ and $a_1(1640),\cdots$,
respectively.

In the $A_z=0$ gauge,
the CS term (\ref{CS}) becomes
\begin{eqnarray}
S_{\rm D8}^{\rm CS}&=&
-\frac{N_c}{24\pi^2}\int_{M^4}\left(
\alpha_4(d\xi_+^{-1}\xi_+,A_L)
-\alpha_4(d\xi_-^{-1}\xi_-,A_R)
\right)\nn\\
&&~~~+\frac{N_c}{24\pi^2}
\int_{M^4\times\bR}\left(\omega_5(A)
-\frac{1}{10}\tr(gdg^{-1})^5\right) \ ,
\label{WZW}
\end{eqnarray}
where $g$ is the gauge function given in (\ref{ginv}),
and $\alpha_4$ reads
\begin{eqnarray}
\alpha_4(V,A)\equiv -\half\tr\left(
V(AdA+dAA+A^3)-\half VAVA-V^3 A
\right) \ .
\end{eqnarray}

The four-dimensional effective action of the mesons
written in terms of $\xi_\pm$ (or $U$) and $B_\mu^{(n)}$,
including the external gauge fields ($A_{L\mu}$, $A_{R\mu}$),
can be obtained by substituting the gauge
potential (\ref{5dpot}) or (\ref{exp})
into the five-dimensional Yang-Mills
action (\ref{DBI}) and the CS-term (\ref{WZW}).
This action is automatically consistent with the symmetry
expressed by (\ref{tr1})-(\ref{tr2}).
The explicit calculation of this effective action is partly given
in Ref.~\citen{SaSu}. It has been shown that the effective action
of the pion is given by the Skyrme model \cite{Skyrme:1} as
\begin{eqnarray}
S_{\rm D8}^{\rm DBI}\Big|_{v^n_\mu=a^n_\mu=\cV_\mu=\cA_\mu=0}
=\int d^4 x\left(\frac{f_\pi^2}{4}\tr\left(U^{-1}\del_\mu U\right)^2+
\frac{1}{32e_S^2}\tr\left[U^{-1}\del_\mu U,U^{-1}\del_\nu U\right]^2
\right) \ ,\nn\\
\label{Skyrme}
\end{eqnarray}
where the pion decay constant $f_\pi$ and 
the dimensionless parameter $e_S$ are given by
\begin{eqnarray}
f_\pi^2&\equiv&\frac{4}{\pi}\kappa=
\frac{1}{54\pi^4}\lambda N_c\ ,
\label{fpi}
\\
e_S^{-2}&\equiv&\kappa \int\! dz\, K^{-1/3}(1-\psi_0^2)^2
\ .
\label{es}
\end{eqnarray}
Also, the CS term (\ref{WZW}) is identical to
the WZW term in QCD that includes the pion field as
well as the external gauge fields
when we omit the vector meson fields $B_\mu^{(n)}$:
\begin{eqnarray}
S^{\rm CS}_{\rm D8}\Big|_{v^n_\mu=a^n_\mu=0}
=-\frac{N_c}{48\pi^2}\int_{M^4}Z
-\frac{N_c}{240\pi^2}\int_{M^4\times \bR}\tr(gdg^{-1})^5 \ ,
\label{WZW0}
\end{eqnarray}
where
\begin{eqnarray}
Z&=&
\tr[(A_R dA_R+dA_R A_R+A_R^3)(U^{-1}A_LU+U^{-1}dU)-{\rm{p.c.}}]\nn\\
&&+\tr[ dA_RdU^{-1}A_L U-{\rm{p.c.}}]
+\tr[A_R(dU^{-1}U)^3-{\rm{p.c.}}]\nn\\
&&+\half\tr[(A_RdU^{-1}U)^2-{\rm{p.c.}}]
+\tr[UA_R U^{-1}A_L dUdU^{-1}-{\rm{p.c.}}]\nn\\
&&-\tr[A_R dU^{-1}UA_R U^{-1}A_LU-{\rm{p.c.}}]
+\half\tr[(A_RU^{-1}A_LU)^2]\ .
\label{Z}
\end{eqnarray}
Here ``p.c.'' represents the terms obtained by
making the exchange $A_L\lra A_R$ and $U\lra U^{-1}$.

In this paper, we analyze the couplings 
among the pions and vector mesons, including the external
gauge fields in more detail. In particular, we examine
whether the vector meson dominance hypothesis
holds for both the DBI part and the WZW term.
We analyze the DBI part in \S \ref{DBIpart}
and the WZW term in \S \ref{WZWterm}.

\section{DBI part}
\label{DBIpart}

\subsection{The effective action}
\label{eff}
In this subsection,
we analyze the effective action obtained by inserting the mode
expansion (\ref{exp}) into the action (\ref{DBI}).
The effective action written in terms of
 $v_\mu^n$, $a_\mu^n$ and $\xi_\pm$,
including the external gauge fields ($A_{L\mu}$, $A_{R\mu}$),
is given in Appendix \ref{AppA}.
Here we consider some of the couplings read off of the action.

It is useful to write the action as
\begin{equation}
S_{\rm D8}^{\rm DBI}=
\int d^4x\,\cL_{\rm div}+\sum_{j\ge 2} \int d^4x \, \cL_j \ ,
\end{equation}
where $\cL_j$ contains 
the terms of order $j$ in the fields
$\Pi,v^n_{\mu},a^n_{\mu},\cV_{\mu}$ and $\cA_{\mu}$.
The quantity $\cL_{\rm div}$ contains the divergent terms that result from
the non-normalizable modes $\cV_\mu$ and $\cA_\mu$.
The explicit form of $\cL_{\rm div}$ is given in (\ref{Ldiv}).

For the quadratic terms, we find
\begin{eqnarray}
\cL_2&=&
\frac{1}{2}\tr\left(\del_{\mu}v_{\nu}^n-\del_{\nu}v_{\mu}^n\right)^2
+\frac{1}{2}\tr\left(\del_{\mu}a_{\nu}^n-\del_{\nu}a_{\mu}^n\right)^2
\nn\\
&&
+a_{\cV v^n}\tr\left(\del^{\mu}\cV^{\nu}-\del^{\nu}\cV^{\mu}\right)
 \left(\del_{\mu}v_{\nu}^n-\del_{\nu}v_{\mu}^n\right)
+a_{\cA a^n}\tr\left(\del^{\mu}\cA^{\nu}-\del^{\nu}\cA^{\mu}\right)
 \left(\del_{\mu}a_{\nu}^n-\del_{\nu}a_{\mu}^n\right) \nn\\
&&
+\tr\left(i\del_{\mu}\Pi+f_{\pi}\cA_{\mu}\right)^2
+m_{v^n}^2\tr\left(v_{\mu}^n\right)^2
+m_{a^n}^2\tr\left(a_{\mu}^n\right)^2
\ ,
\label{L2}
\end{eqnarray}
where
\begin{eqnarray}
m_{v^n}^2\equiv\lambda_{2n-1}\ ,
~~~m_{a^n}^2\equiv\lambda_{2n}\ ,
\label{m2lam}
\end{eqnarray}
\begin{eqnarray}
a_{\cV v^n}\equiv\kappa\int\! dz\, K^{-1/3}\psi_{2n-1}\ ,~~~
a_{\cA a^n}\equiv\kappa\int\! dz\, K^{-1/3}\psi_{2n}\psi_0\ ,
\label{aVvaAa}
\end{eqnarray}
and we have used the fact that the pion decay constant
$f_\pi$ is given by (\ref{fpi}).
Here and in the following, the summation symbol
``$\sum_{n=1}^\infty$'' is often
omitted for notational simplicity.

In order to diagonalize the kinetic term, we define
\begin{eqnarray}
\wt v_\mu^n&\equiv& v_\mu^n+a_{\cV v^n}\cV_\mu \ ,
\label{wtv}\\
\wt a_\mu^n&\equiv& a_\mu^n+a_{\cA a^n}\cA_\mu\ .
\label{wta}
\end{eqnarray}
Then, (\ref{L2}) becomes
\begin{eqnarray}
\cL_2&=&
\frac{1}{2}\tr\left(
\del_{\mu}\wt v_{\nu}^n-\del_{\nu}\wt v_{\mu}^n\right)^2
+\frac{1}{2}\tr\left(
\del_{\mu}\wt a_{\nu}^n-\del_{\nu}\wt a_{\mu}^n\right)^2
+\tr\left(i\del_{\mu}\Pi+f_{\pi}\cA_{\mu}\right)^2
\nn\\
&&
+m_{v^n}^2\tr\left(\wt v_{\mu}^n-a_{\cV v^n}\cV_\mu
\right)^2
+m_{a^n}^2\tr\left(\wt a_{\mu}^n-a_{\cA a^n}\cA_\mu
\right)^2
\ .
\end{eqnarray}
Here, corrections to the kinetic terms of $\cV_\mu$ and $\cA_\mu$
in $\cL_{\rm div}$ are omitted.

We segregate the cubic terms $\cL_3$ into terms of equal orders in
the pion field $\Pi(x^\mu)$:
\begin{eqnarray}
\cL_3=\cL_3|_{\pi^0}+\cL_3|_{\pi^1}+\cL_3|_{\pi^2}\ .
\end{eqnarray}
Note that $\cL_3|_{\pi^3}$ does not exist because of
parity symmetry.

Let us first examine $\cL_3|_{\pi^2}$, which is relevant
to the electromagnetic form factor of the pion:
\begin{eqnarray}
\cL_3|_{\pi^2}&=&
\frac{b_{\cV\pi\pi}}{f_\pi^2}
\tr\left(
(\del_\mu\cV_\nu-\del_\nu\cV_\mu)[\del^\mu\Pi,\del^\nu\Pi\,]
\right)\nn\\
&&+\frac{b_{v^n\pi\pi}}{f_\pi^2}
\tr\left(
(\del_\mu v^n_\nu-\del_\nu v^n_\mu)[\del^\mu\Pi,\del^\nu\Pi\,]
\right)
-2\tr\left(\cV_\mu[\,\Pi,\del^\mu\Pi\,]\right)\\
&=&
\frac{1}{f_\pi^2}\left(
b_{\cV\pi\pi}-a_{\cV v^n}b_{v^n\pi\pi}\right)
\tr\left(
(\del_\mu\cV_\nu-\del_\nu\cV_\mu)[\del^\mu\Pi,\del^\nu\Pi\,]
\right)\nn\\
&&+\frac{b_{v^n\pi\pi}}{f_\pi^2}
\tr\left(
(\del_\mu\wt v^n_\nu-\del_\nu\wt v^n_\mu)[\del^\mu\Pi,\del^\nu\Pi\,]
\right)
-2\tr\left(\cV_\mu[\,\Pi,\del^\mu\Pi\,]\right)\ ,
\label{L3_2}
\end{eqnarray}
where
\begin{eqnarray}
b_{\cV \pi\pi}&\equiv&\kappa\int\! dz\, K^{-1/3}
\left(1-\psi_0^2\right)\ ,
\label{cVpp}\\
b_{v^n\pi\pi}&\equiv&\kappa\int\! dz\, K^{-1/3}
\psi_{2n-1}\left(1-\psi_0^2\right)\ .
\label{vnpp}
\end{eqnarray}
Note here that the coefficient of the
first term in (\ref{L3_2}) is zero.
Actually, using the completeness relation
\begin{eqnarray}
\kappa\sum_{n=1}^\infty K^{-1/3}(z')\psi_n(z)\psi_n(z')
=\delta(z-z')\ ,
\label{comp}
\end{eqnarray}
we can verify that
\begin{eqnarray}
\sum_{n=1}^\infty a_{\cV v^n}b_{v^n\pi\pi}
= b_{\cV\pi\pi}\ .
\label{sum}
\end{eqnarray}
Therefore, (\ref{L3_2}) becomes
\begin{eqnarray}
\cL_3|_{\pi^2}&=&
\frac{b_{v^n\pi\pi}}{f_\pi^2}
\tr\left(
(\del_\mu\wt v^n_\nu-\del_\nu\wt v^n_\mu)[\del^\mu\Pi,\del^\nu\Pi\,]
\right)
-2\tr\left(\cV_\mu[\,\Pi,\del^\mu\Pi\,]\right)\ .
\end{eqnarray}

In order to compare the effective action
with that given in the literature (e.g. Ref.~\citen{bky;review})
we rewrite the Lagrangian using
\begin{eqnarray}
\wh v_\mu^n\equiv\wt v_\mu^n+\frac{b_{v^n\pi\pi}}{2f_\pi^2}
\,[\Pi,\del_\mu\Pi]\ ,
\label{whv}
\end{eqnarray}
and remove the term of the form
$\tr\left((\del_\mu\wt v^n_\nu-\del_\nu \wt v^n_\mu)
[\del^\mu\Pi,\del^\nu\Pi]\right)$.
Then, we obtain
\begin{eqnarray}
\cL_2+\cL_3|_{\pi^2}&=&
\frac{1}{2}\tr\left(
\del_{\mu}\wh v_{\nu}^n-\del_{\nu}\wh v_{\mu}^n\right)^2
+\frac{1}{2}\tr\left(
\del_{\mu}\wt a_{\nu}^n-\del_{\nu}\wt a_{\mu}^n\right)^2
+\tr\left(i\del_{\mu}\Pi+f_{\pi}\cA_{\mu}\right)^2\nn\\
&&
+m_{a^n}^2\tr\left(\wt a_{\mu}^n-a_{\cA a^n}\cA_\mu\right)^2
+m_{v^n}^2\tr\left(\wh v_{\mu}^n
-a_{\cV v^n}\cV_\mu\right)^2\nn\\
&&
-\frac{m_{v^n}^2b_{v^n\pi\pi}}{f_\pi^2}
\tr\left(\wh v_\mu^n[\Pi,\del^\mu\Pi]\right)
+\left(\frac{m_{v^n}^2b_{v^n\pi\pi}a_{\cV v^n}}{f_\pi^2}-2\right)
\tr\left(\cV_\mu[\Pi,\del^\mu\Pi]\right)\nn\\
&&
+\frac{m_{v^n}^2b_{v^n\pi\pi}^2}{4f_{\pi}^4}
\tr\left[ \Pi,\del_{\mu}\Pi\right]^2
-\frac{b_{v^n\pi\pi}^2}{2f_{\pi}^4}
\tr\left[ \del_{\mu}\Pi,\del_{\nu}\Pi\right]^2
\ .
\label{L2L3}
\end{eqnarray}
Here, it is very important to note the relation
\begin{eqnarray}
&&\sum_{n=1}^\infty
m_{v^n}^2 b_{v^n\pi\pi}a_{\cV v^n} =2f_{\pi}^2 \ ,
\label{VMD}
\end{eqnarray}
which follows straightforwardly from the completeness condition
(\ref{comp}) and the equation (\ref{deqn;psi}).
This shows that the $\pi\pi\cV$ coupling in (\ref{L2L3}) vanishes.
As shown in \S \ref{s:VMD},
this fact is important with regard to
the vector meson dominance in the electromagnetic form factor
of the pion. Similarly,
we can also show the following relations:\footnote{
The sum rules (\ref{sum}), (\ref{VMD}) and (\ref{coe:pi4})
for a closely related five-dimensional model
are also derived in Ref.~\citen{Hirn:2005nr} using a similar method.
}
\begin{equation}
\sum_{n=1}^\infty m_{v^n}^2b_{v^{n}\pi\pi}^2=\frac{4}{3}f_{\pi}^2 \ ,
~~~
\sum_{n=1}^\infty b_{v^{n}\pi\pi}^2=
e_S^{-2} \ .
\label{coe:pi4}
\end{equation}
Using (\ref{VMD}) and (\ref{coe:pi4}),
the Lagrangian (\ref{L2L3}) can be rewritten as
\begin{eqnarray}
\cL_2+\cL_3|_{\pi^2}&=&
\frac{1}{2}\tr\left(
\del_{\mu}\wh v_{\nu}^n-\del_{\nu}\wh v_{\mu}^n\right)^2
+\frac{1}{2}\tr\left(
\del_{\mu}\wt a_{\nu}^n-\del_{\nu}\wt a_{\mu}^n\right)^2
+\tr\left(i\del_{\mu}\Pi+f_{\pi}\cA_{\mu}\right)^2\nn\\
&&
+m_{a^n}^2\tr\left(\wt a_{\mu}^n\right)^2
-2 g_{a^n}\tr\left(\wt a_{\mu}^n\cA^\mu\right)
+m_{a^n}^2a_{\cA a^n}^2
\tr\left(\cA_{\mu}\right)^2\nn\\
&&
+m_{v^n}^2\tr\left(\wh v_{\mu}^n\right)^2
-2 g_{v^n}\tr\left(\wh v_{\mu}^n\cV^\mu\right)
+m_{v^n}^2a_{\cV v^n}^2
\tr\left(\cV_{\mu}\right)^2\nn\\
&&
-2 g_{v^n\pi\pi}\tr\left(\wh v_\mu^n[\Pi,\del^\mu\Pi]\right)\nn\\
&&
+\frac{1}{3f_\pi^2}\tr\left[ \Pi,\del_{\mu}\Pi\right]^2
-\frac{1}{2e_S^2f_{\pi}^4 }
\tr\left[ \del_{\mu}\Pi,\del_{\nu}\Pi\right]^2
\ ,
\label{L2L3_2}
\end{eqnarray}
where
\begin{eqnarray}
g_{a^n}\equiv m_{a^n}^2a_{\cA a^n}\ ,~~~
g_{v^n}\equiv m_{v^n}^2 a_{\cV v^n}\ ,~~~
g_{v^n\pi\pi}\equiv\frac{b_{v^n\pi\pi}m_{v^n}^2}{2f_\pi^2}\ .
\label{couplings}
\end{eqnarray}
Here, we can verify that $g_{v^n}$ and $g_{a^n}$ are
equal to the decay constants
of the vector meson $v^n$ and the axial-vector meson $a^n$, respectively,
by showing that
\begin{eqnarray}
\langle 0|J_{\mu}^{(V)a}(0)|v^{nb}\rangle=
g_{v^n}\delta^{ab}\epsilon_{\mu} \ , ~~~
\langle 0|J_{\mu}^{(A)a}(0)|a^{nb}\rangle=
g_{a^n}\delta^{ab}\epsilon_{\mu} \ , 
\end{eqnarray}
where the quantities $J_{\mu}^{(V,A)}$ are the conserved
vector and axial-vector currents coupled with $\cV^{\mu}$ and $\cA^{\mu}$,
respectively, $\epsilon_\mu$ are the polarizations of the
vector mesons, and the indices $a$ and $b$
 are associated with the generators $T^a$
of $U(N_f)$ as
\begin{equation}
v_{\mu}^{n}=iv_{\mu}^{na}T^a \ ,~~
a_{\mu}^{n}=ia_{\mu}^{na}T^a \ ,~~
\cV_{\mu}=i\cV_{\mu}^{a}T^a \ ,~~
\cA_{\mu}=i\cA_{\mu}^{a}T^a \ .
\end{equation}
Note that the decay constants can be recast as
\begin{eqnarray}
g_{v^n}&\!\!=\!\!&-\kappa\int\! dz\,
 \del_z\left(K\del_z\psi_{2n-1}\right)
=-2\kappa(K\del_z\psi_{2n-1})\Big|_{z=+\infty} \ ,\nn\\
g_{a^n}&\!\!=\!\!&
-\kappa\int\! dz\, \psi_0\,\del_z\left(K\del_z\psi_{2n}\right) 
=-2\kappa(K\del_z\psi_{2n})\Big|_{z=+\infty} \ ,
\label{dconst2}
\end{eqnarray}
where we have used (\ref{deqn;psi}).
This shows that the decay constants $g_{v^n}$ and $g_{a^n}$
are fixed uniquely by the
asymptotic behavior of the mode functions $\psi_{2n-1}$
and $\psi_{2n}$, respectively.
(See Ref.~\citen{SoSt} for analogous formulas.)

The terms linear in $\Pi$ are
\begin{eqnarray}
\cL_3|_{\pi^1}&=&\frac{2i}{f_\pi}\tr\Bigg[
\del^\mu(\del_\mu\cV_\nu-\del_\nu\cV_\mu)[\Pi,\cA^\nu]\, b_{\cV\pi\pi}+
\del^\mu(\del_\mu v^n_\nu-\del_\nu v^n_\mu)[\Pi,\cA^\nu]\, b_{v^n\pi\pi}
\nn\\
&&+
\del^\mu(\del_\mu\cA_\nu-\del_\nu\cA_\mu)[\Pi,v^{n\nu}]\, 
(b_{v^n\pi\pi}-a_{\cV v^n})\nn\\
&&+\del^\mu(\del_\mu\cV_\nu-\del_\nu\cV_\mu)[\Pi,a^{m\nu}]\, 
(-a_{\cA a^m})+
\del^\mu(\del_\mu v^n_\nu-\del_\nu v^n_\mu)[\Pi,a^{m\nu}]\, 
(-c_{v^n a^m\pi})\nn\\
&&+
\del^\mu(\del_\mu a^m_\nu-\del_\nu a^m_\mu)[\Pi,v^{n\nu}]\, 
(-c_{v^n a^m\pi})\Bigg]\nn\\
&&-2if_\pi\tr\left(\cA_\mu[\Pi,\cV^\mu]\right)\ ,
\label{pi1}
\end{eqnarray}
up to total derivative terms.
Here, we have defined
\begin{eqnarray}
c_{v^n a^m\pi}\equiv 
\kappa\int\! dz\, K^{-1/3}\psi_0\psi_{2n-1}\psi_{2m}\ .
\label{cpva}
\end{eqnarray}
Using the sum rule (\ref{sum}) and the relations
\begin{eqnarray}
\sum_{n=1}^\infty a_{\cV v^n}c_{v^n a^m\pi}&=&a_{\cA a^m}\ ,
\label{aca}\\
\sum_{m=1}^\infty a_{\cA a^m}c_{v^n a^m\pi}
&=&a_{\cV v^n}-b_{v^n\pi\pi}\ ,
\end{eqnarray}
which also follow from the completeness condition (\ref{comp}),
the Lagrangian (\ref{pi1}) can be rewritten as
\begin{eqnarray}
\cL_3|_{\pi^1}&=&\frac{2i}{f_\pi}\tr\Bigg[
\del^\mu(\del_\mu\wh v^n_\nu-\del_\nu\wh v^n_\mu)
[\Pi,\cA^\nu]\, a_{\cV v^n}+
\del^\mu(\del_\mu\wt a^m_\nu-\del_\nu\wt a^m_\mu)
[\Pi,\cV^\nu]\, a_{\cA a^m}\nn\\
&&+
\del^\mu(\del_\mu \wh v^n_\nu-\del_\nu \wh v^n_\mu)
[\Pi,\wt a^{m\nu}]\, (-c_{v^n a^m\pi})+
\del^\mu(\del_\mu \wt a^m_\nu-\del_\nu \wt a^m_\mu)
[\Pi,\wh v^{n\nu}]\, (-c_{v^n a^m\pi})\Bigg]\nn\\
&&-2if_\pi\tr\left(\cA_\mu[\Pi,\cV^\mu]\right)
+\mbox{(quartic terms)}\ .
\label{pi1_2}
\end{eqnarray}

Finally, we consider the rest of $\cL_3$, which contains no
pion field $\Pi(x^\mu)$.
It can be shown that
\begin{eqnarray}
&&\cL_3|_{\pi^0}\nn\\
&=&\tr\Bigg(
(\del^\mu\cV^\nu-\del^\nu\cV^\mu)\nn\\
&&~~\times\Big\{
([\cV_\mu,v^n_\nu]-[\cV_\nu,v^n_\mu])\, a_{\cV v^n}+
([\cA_\mu,a^n_\nu]-[\cA_\nu,a^n_\mu])\, a_{\cA a^n}+
[v^n_\mu,v_\nu^n]+[a^n_\nu,a^n_\mu]\Big\}\nn\\
&&~+(\del^\mu v^{l\nu}-\del^\nu v^{l\mu})
\Big\{[\cV_\mu,\cV_\nu]\, a_{\cV v^l}+
[\cA_\mu,\cA_\nu](a_{\cV v^l}-b_{v^l\pi\pi})+
([\cV_\mu,v^l_\nu]-[\cV_\nu,v^l_\mu])\nn\\
&&~~~~~+
([\cA_\mu,a^m_\nu]-[\cA_\nu,a^m_\mu])\, c_{v^n a^m\pi}+
[v^m_\mu,v^n_\nu]\, g_{v^l v^m v^n}+
[a^m_\mu,a^n_\nu]\, g_{v^l a^m a^n}\Big\}\nn\\
&&~+(\del^\mu \cA^\nu-\del^\nu \cA^\mu)
\Big\{([\cV_\mu,a^n_\nu]-[\cV_\nu,a^n_\mu])\, a_{\cA a^n}\nn\\
&&~~~~~+
([\cA_\mu,v^n_\nu]-[\cA_\nu,v^n_\mu])(a_{\cV v^n}-b_{v^n\pi\pi})+
([v^n_\mu,a^m_\nu]-[v^n_\nu,a^m_\mu])\,c_{v^n a^m\pi}\Big\}\nn\\
&&~+(\del^\mu a^{n\nu}-\del^\nu a^{n\mu})
\Big\{
([\cV_\mu,\cA_\nu]-[\cV_\nu,\cA_\mu])\, a_{\cA a^n}+
([\cV_\mu,a^n_\nu]-[\cV_\nu,a^n_\mu])\nn\\
&&~~~~~+
([\cA_\mu,v^m_\nu]-[\cA_\nu,v^m_\mu])\,c_{v^m a^n\pi}+
([v^l_\mu,a^m_\nu]-[v^l_\nu,a^m_\mu])\,g_{v^l a^m a^n}\Big\}\Bigg)
\ ,
\label{L3pi0}
\end{eqnarray}
where
\begin{eqnarray}
g_{v^lv^mv^n}&\!\equiv\!&
\kappa\int\! dz\, K^{-1/3}\psi_{2l-1}\psi_{2m-1}\psi_{2n-1} \ ,\nn \\ 
g_{v^la^ma^n}&\!\equiv\!&
\kappa\int\! dz\, K^{-1/3}\psi_{2l-1}\psi_{2m}\psi_{2n} \ .
\end{eqnarray}
If we rewrite (\ref{L3pi0}) in terms of $\wt v^n_\mu$ and  $\wt a^n_\mu$
defined in (\ref{wtv}) and (\ref{wta}), we obtain
\begin{eqnarray}
\cL_3|_{\pi^0}&\!=\!&
\tr\Bigg(
g_{v^lv^mv^n}
\left(\del^{\mu}\wt v^{l \nu}-\del^{\nu}\wt v^{l \mu}\right)
\left[ \wt v^m_{\mu}, \wt v^n_{\nu}\right]
+
g_{v^la^ma^n}
\left(\del^{\mu}\wt v^{l \nu}-\del^{\nu}\wt v^{l \mu}\right)
\left[ \wt a^m_{\mu}, \wt a^n_{\nu}\right]\nn\\
&&+
g_{v^la^ma^n}
\Big(\del^{\mu}\wt a^{n \nu}-\del^{\nu}\wt a^{n \mu}\Big)
\Big\{
\left[ \wt v^l_{\mu}, \wt a^m_{\nu}\right]
-\left[ \wt v^l_{\nu}, \wt a^m_{\mu}\right]
\Big\}
\Bigg)\ .
\label{vvv}
\end{eqnarray}
Here, corrections to the cubic term in $\cL_{\rm div}$
are omitted. From (\ref{vvv}), we see that
the direct cubic couplings of
the vector mesons $v^n_\mu$ and $a^n_\mu$ to
the external gauge fields $\cV_\mu$ and $\cA_\mu$
disappear.

For $\cL_4$, we focus on the quartic terms in the pion field.
As explained in \S \ref{rev},
the low energy effective action of the pion
is given by the Skyrme model (\ref{Skyrme}).
Then, it follows from (\ref{Skyrme}) that
\begin{equation}
\cL_4|_{\pi^4}=-\frac{1}{3f_{\pi}^2}
\tr\left[ \Pi,\del_{\mu}\Pi\right]^2
+\frac{1}{2e_S^2f_{\pi}^4}
\tr\left[ \del_{\mu}\Pi,\del_{\nu}\Pi\right]^2\ .
\label{L4}
\end{equation}
Note that (\ref{L4}) exactly cancels the $\cO(\Pi^4)$ terms
(the last two terms) in (\ref{L2L3_2}).

\subsection{Numerical results}
\label{numerical}

Here, we summarize the numerically obtained
values of the coupling constants
to provide a rough estimate of the physical quantities.
Listed below are the numerical estimates for
some of the masses and coupling constants
defined in (\ref{m2lam}) and (\ref{couplings})
in units for which $\Mkk=1$.
\begin{eqnarray}
\begin{array}{c||ccc|cc}
n& m_{v^n}^2&\kappa^{-1/2}g_{v^n}
&\kappa^{1/2} g_{v^n\pi\pi}&m_{a^n}^2&\kappa^{-1/2} g_{a^n}\\
\hline
1&0.669& 2.11 & 0.415&1.57&5.02\\
2&2.87 & 9.10 & -0.109&4.55&14.4\\
3&6.59 & 20.8&0.0160&9.01&28.3\\
4&11.8 & 37.1&-0.00408&15.0&46.9
\end{array}
\label{table}
\end{eqnarray}
These are obtained by
solving the equation (\ref{deqn;psi}) numerically
using the shooting method, as in Ref.~\citen{SaSu}.

The coupling constants given in
(\ref{fpi}), (\ref{es}) and (\ref{cVpp}) are
easily calculated as
\begin{eqnarray}
f_\pi^2
\simeq 1.27\cdot\kappa\ ,~~~
e_S^{-2}\simeq 2.51\cdot\kappa\ ,~~~
b_{\cV\pi\pi}\simeq 4.69\cdot\kappa\ . 
\label{feb}
\end{eqnarray}

It is,  however, important to keep in mind that we should not
take these numerical values too seriously, because
the approximation made in our analysis is very crude.
As discussed in Ref.~\citen{SaSu},
the present model deviates from realistic QCD above
the energy scale of $\Mkk$, which is the same
as the mass scale of the vector mesons.
Furthermore, all the quarks are assumed to be massless,
and the supergravity description and the probe
approximation are valid only when $N_c\gg N_f$
and $\lambda\gg 1$.
 In the following subsections,
we compare the numerical values of the coupling constants
obtained in our model with the experimental values 
in order to get some idea of whether or not we are on the right track.
It would be interesting to
improve the approximation in order to make more accurate predictions.

\subsection{The Skyrme term}
The second term in (\ref{Skyrme}), which is called the Skyrme term,
can be written as
\begin{eqnarray}
\frac{1}{32e_S^2}\tr\left[U^{-1}\del_\mu U,U^{-1}\del_\nu U\right]^2
=L_1 P_1+L_2 P_2+L_3 P_3
\end{eqnarray}
for $N_f=3$, where
\begin{eqnarray}
&&P_1\equiv \left[\tr(\del_\mu U^{-1}\del^\mu U)\right]^2\ ,~~~
P_2\equiv
\tr(\del_\mu U^{-1}\del_\nu U)
\tr(\del^\mu U^{-1}\del^\nu U)\ ,\nn\\
&&P_3\equiv
\tr(\del_\mu U^{-1}\del^\mu U\del_\nu U^{-1}\del^\nu U) \ ,
\end{eqnarray}
and
\begin{eqnarray}
L_1=\frac{1}{32 e_S^2}\ ,~~~
L_2=\frac{1}{16 e_S^2}\ ,~~~
L_3=-\frac{3}{16 e_S^2}\ .
\label{L123}
\end{eqnarray}
For the case $N_f=2$,
we have the additional relation $P_3=\half P_1$.
The experimental values for the coefficients $L_i$ ($i=1,2,3$)
(at the scale of the $\rho$ meson mass)
are given in Ref.~\citen{Pich} as
\begin{eqnarray}
L_1|_{\rm exp}&\simeq& (0.4\pm 0.3)\times 10^{-3},\nn\\
L_2|_{\rm exp}&\simeq& (1.4\pm 0.3)\times 10^{-3},\nn\\
L_3|_{\rm exp}&\simeq& (-3.5\pm 1.1)\times 10^{-3}.
\end{eqnarray}
Our result (\ref{L123}) is roughly
consistent with experimental results
in the case $\kappa\simeq (7-9)\times 10^{-3}$.
Note that this value of $\kappa$ is also consistent
with that obtained by combining the experimental
values $f_\pi|_{\rm exp}\simeq 92.4~{\rm MeV}$ and
$m_{\rho}|_{\rm exp}\simeq 776~{\rm MeV}$ with
the numerical results (\ref{table}) and (\ref{feb}):
\begin{eqnarray}
\Mkk\simeq 949~{\rm MeV}\ ,~~~
\kappa\simeq 7.45 \times 10^{-3}\ .
\label{kappa}
\end{eqnarray}

\subsection{KSRF relations}
\label{KSRF}

Here we examine the KSRF relations \cite{KS,RF},
\begin{eqnarray}
g_{\rho} &=& 2g_{\rho\pi\pi}f_\pi^2~~~~~ \mbox{(KSRF(I))}\ ,\\
m_{\rho}^2 &=& 2g_{\rho\pi\pi}^2f_\pi^2~~~~~ \mbox{(KSRF(II))}
\label{KSRF2}\ .
\end{eqnarray}
These two relations leads to
\begin{eqnarray}
g_{\rho} g_{\rho\pi\pi}= m_\rho^2\ .
\label{ggm}
\end{eqnarray}
Then, using the experimental values
$g_{\rho\pi\pi}|_{\rm exp}\simeq 5.99$ and
$g_{\rho}|_{\rm exp}\simeq 0.121~{\rm GeV}^2$, \cite{pdg}
we obtain
\begin{eqnarray}
\left.\frac{4 g_{\rho\pi\pi}^2 f_\pi^2}{m_\rho^2}
\right|_{\rm exp}\simeq 2.03\ ,~~~
\left.\frac{g_\rho g_{\rho\pi\pi}}{m_\rho^2}
\right|_{\rm exp}\simeq 1.20\ ,
\end{eqnarray}
which show that the relations (\ref{KSRF2}) and (\ref{ggm})
are satisfied to within 20\%.

The corresponding values in our model can be estimated
by using the numerical values listed in \S \ref{numerical}.
The result is
\begin{eqnarray}
\frac{4 g_{v^1\pi\pi}^2 f_\pi^2}{m_{v^1}^2}
\simeq 1.31\ ,~~~
\frac{g_{v^1} g_{v^1\pi\pi}}{m_{v^1}^2}
\simeq 1.31\ .
\label{131}
\end{eqnarray}
Note that these values are independent of the parameters in the model.
If we use the values of $\Mkk$ and $\kappa$ in (\ref{kappa}),
we obtain
\begin{eqnarray}
g_{v^1\pi\pi}\simeq 4.81\ ,
~~~ g_{v^1}\simeq 0.164~{\rm GeV}^2\ .
\end{eqnarray}

Remarkably, it is found that the relations given in
(\ref{131}) are equivalent
to picking out the dominant contribution from the following
sum rules:
\begin{eqnarray}
\sum_{n=1}^\infty
\frac{4 g_{v^n\pi\pi}^2f_\pi^2}{m_{v^n}^2}&=&\frac{4}{3} \ ,
\label{sum1}\\
\sum_{n=1}^\infty
\frac{g_{v^n}g_{v^n\pi\pi}}{m_{v^n}^2}&=&1 \ .
\label{VMD:2}
\end{eqnarray}
The relation (\ref{sum1}) is equivalent to the first relation
in (\ref{coe:pi4}), and (\ref{VMD:2}) follows from (\ref{VMD}).
The sum rules given in (\ref{sum1}) and (\ref{VMD:2}) were
first reported in Refs.~\citen{DaRoPo} and \citen{SoSt},
respectively, and have been
shown to be satisfied in general five-dimensional models.
As a check, using the numerical results for $n=1,2,3$ and $4$
given in (\ref{table}),
the left-hand sides of (\ref{sum1}) and (\ref{VMD:2})
are evaluated as
\begin{eqnarray}
\sum_{n=1}^\infty
\frac{4 g_{v^n\pi\pi}^2f_\pi^2}{m_{v^n}^2}
&\simeq&1.31+0.0210+0.000197+0.00000717+
\cdots\simeq 1.33\ ,\\
\sum_{n=1}^\infty\frac{g_{v^n}g_{v^n\pi\pi}}{m_{v^n}^2}
&\simeq&1.31-0.346+0.0505-0.0128+\cdots\simeq 1.00\ ,
\end{eqnarray}
{}from which we see that the contribution of the lightest
vector meson $\wh v^1$ (the $\rho$ meson) dominates the sum.

\subsection{Electromagnetic form factors}
\label{s:VMD}

Let us consider the pion form factor $F_\pi(p^2)$
defined by
\begin{eqnarray}
\langle{\pi^a(p)}|J_{\mu}^{(V)c}(0)|
\pi^b(p')\rangle=f^{abc}
(p+p')_\mu F_\pi((p-p')^2)\ ,
\end{eqnarray}
where $f^{abc}$ is the structure constant of $U(N_f)$.

\begin{figure}
\begin{center}
\begin{picture}(100,65)(0,-5)
\put(0,0){
\includegraphics[scale=0.20]{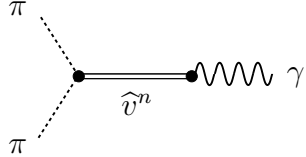}}
\put(-3,50){\makebox(0,0){$\pi$}}
\put(-3,-2){\makebox(0,0){$\pi$}}
\put(42,14){\makebox(0,0){$\wh v^n$}}
\put(102,24){\makebox(0,0){$\gamma$}}
\end{picture}
\caption{\footnotesize{
Pion form factor.
}}
\label{fig1}
\end{center}
\end{figure}

Combining the $\wh v^n\cV$ and $\wh v^n\pi\pi$
vertices in (\ref{L2L3_2}) as well as the $\wh v^n$ propagators,
as depicted in Fig. \ref{fig1},
we obtain
\begin{equation}
F_{\pi}(k^2)=\sum_{n=1}^\infty
\frac{g_{v^n}g_{v^n\pi\pi}}{k^2+m_{v^n}^2} \ .
\label{form:pi}
\end{equation}
Crucial in this computation is the relation (\ref{VMD}), which ensures
that the direct $\cV\pi\pi$ coupling in (\ref{L2L3}) vanishes.
As a consistency check of (\ref{form:pi}), we note that
\begin{equation}
F_{\pi}(0)=
\sum_{n=1}^\infty\frac{g_{v^n}g_{v^n\pi\pi}}{m_{v^n}^2}=1 \ ,
\label{F0}
\end{equation}
due to (\ref{VMD:2}).

We thus see that our model possesses vector meson dominance 
for the pion form factor (\ref{form:pi}), because the form factor is
saturated by the exchange of vector mesons.
This fact was first pointed out in Ref.~\citen{SoSt}, in which
(\ref{form:pi}) is derived by taking the continuum limit
in the discretized version of the five-dimensional model.
General analyses of the vector meson dominance
in various holographic models are also given in
Refs.~\citen{HoYoSt1,HoYoSt2} and \citen{DaRoPo}.
As we have seen in \S \ref{KSRF},
the sum (\ref{VMD:2}) is dominated by the $\rho$ meson.
Hence, our model exhibits $\rho$ meson dominance
to a good approximation in the form factor $F_{\pi}(k^2)$.
Manifestation of the vector meson dominance in the WZW term
is examined in the next section, and more general consideration
is given in \S \ref{revisit}.

{}By expanding the form factor in $k^2$ as
\begin{equation}
F_{\pi}(k^2)=1-\sum_{n=1}^\infty
\frac{g_{v^n}g_{v^n\pi\pi}}{m_{v^n}^4}k^2
+\cO(k^4) \ ,
\end{equation}
we can extract the charge radius of the pion as
\begin{eqnarray}
\langle r^2 \rangle^{\pi^\pm}= 6
\sum_{n=1}^\infty\frac{g_{v^n}g_{v^n\pi\pi}}{m_{v^n}^4}\ .
\end{eqnarray}
Using the sum rule (\ref{sum}), we can show
\begin{eqnarray}
\sum_{n=1}^\infty\frac{g_{v^n}g_{v^n\pi\pi}}{m_{v^n}^4}
=\frac{1}{2 f_\pi^2}
\sum_{n=1}^\infty a_{\cV v^n} b_{v^n\pi\pi}
=\frac{\pi}{8}\kappa^{-1} b_{\cV\pi\pi}\simeq 1.84 \cdot \Mkk^{-2}\ ,
\end{eqnarray}
and hence
\begin{eqnarray}
\langle r^2 \rangle^{\pi^\pm}
=\frac{3\pi}{4}\kappa^{-1}b_{\cV\pi\pi}\simeq 11.0\cdot\Mkk^{-2}\ ,
\end{eqnarray}
where we have recovered $\Mkk$ in the last expression.
If we use the value of $\Mkk$ given
in (\ref{kappa}), we have
\begin{eqnarray}
\langle r^2 \rangle^{\pi^\pm} \simeq (0.690~{\rm fm})^2\ .
\end{eqnarray}
The experimental value for this is \cite{pdg}
\begin{eqnarray}
\langle r^2 \rangle^{\pi^\pm}\Big|_{\rm exp}
\simeq (0.672~{\rm fm})^2\ .
\end{eqnarray}

It is also interesting to note the sum rules
\begin{eqnarray}
\sum_{n=1}^\infty
\frac{g_{v^n}g_{v^nv^kv^l}}{m_{v^n}^2}=\delta_{kl} \ ,
~~~\sum_{n=1}^\infty
\frac{g_{v^n}g_{v^na^ka^l}}{m_{v^n}^2}=\delta_{kl} \ ,
\label{vasum}
\end{eqnarray}
which are the analogs of (\ref{F0}) for (axial-)vector mesons.
If the sums in both (\ref{F0}) and (\ref{vasum}) (for $k=l$)
are dominated by the contribution of the $\rho$ meson ($n=1$),
we obtain the approximate relation
\begin{eqnarray}
g_{\rho\pi\pi}\simeq g_{\rho v^mv^m}\simeq g_{\rho a^ma^m}
\simeq\frac{m_{\rho}^2}{g_\rho}\ ,
\end{eqnarray}
which leads to the universality
of the $\rho$ meson couplings,
\begin{equation}
g_{\rho HH}\simeq\frac{m_{\rho}^2}{g_{\rho}} \ , ~~~
(H=\pi,v^m,a^m)
\label{univ}
\end{equation}
as discussed in Ref.~\citen{HoYoSt2}.

In order to determine the extent to which
the relation (\ref{univ}) is valid,
we list some numerical results for $g_{\rho v^nv^n}$
and $g_{\rho a^na^n}$:
\begin{eqnarray}
\begin{array}{c||cc}
n & \kappa^{1/2}g_{\rho v^nv^n} &\kappa^{1/2}g_{\rho a^na^n} \\
\hline
1 & 0.447 & 0.286 \\
2 & 0.269 & 0.257 \\
3 & 0.252 & 0.249 \\
4 & 0.247 & 0.246
\end{array}
\label{table1}
\end{eqnarray}
As argued in Ref.~\citen{SaSu},
$g_{\rho \pi\pi}$ and $g_{\rho\rho\rho }$
are nearly equal.
However, these two values are not in good agreement with those of
$g_{\rho v^nv^n}~ (n\ge 2)$ and 
$g_{\rho v^nv^n}~ (n\ge 1)$, among which the universality holds
to a good approximation.
The contributions from the first five terms in the
summations in (\ref{vasum}) for $k=l=1,2$ are estimated as
\begin{eqnarray}
\sum_{n=1}^\infty\frac{g_{v^n}g_{v^n v^1v^1}}{m_{v^n}^2}
&\simeq&1.41-0.464+0.0581-0.00116+0.000845+\cdots\simeq 1.00\ ,\\
\sum_{n=1}^\infty\frac{g_{v^n}g_{v^n v^2v^2}}{m_{v^n}^2}
&\simeq&0.846+0.135+0.381-0.466+0.0993+\cdots\simeq 0.995\ ,\\
\sum_{n=1}^\infty\frac{g_{v^n}g_{v^n a^1a^1}}{m_{v^n}^2}
&\simeq&0.902+0.467-0.453-0.0822+0.00273+\cdots\simeq 1.00\ ,\\
\sum_{n=1}^\infty\frac{g_{v^n}g_{v^n a^2a^2}}{m_{v^n}^2}
&\simeq& 0.810+0.119+0.104+0.316-0.468+\cdots\simeq 0.882\ .
\end{eqnarray}

\subsection{$a_1\ra \pi\gamma$ and $a_1\ra \pi\rho$ decay}

The $a_1$ meson is the lightest axial-vector meson of
$J^{PC}=1^{++}$. In our model, it is
identified with $B_{\mu}^{(n=2)}=a_{\mu}^1$.
Here we discuss the decay amplitudes of $a_1\ra \pi\gamma$
and $a_1\ra\pi\rho$.

\begin{figure}
\begin{center}
\begin{picture}(100,80)(0,-8)
\put(-53,50){\makebox(0,0){$(1)$}}
\put(-50,0){
\includegraphics[scale=0.17]{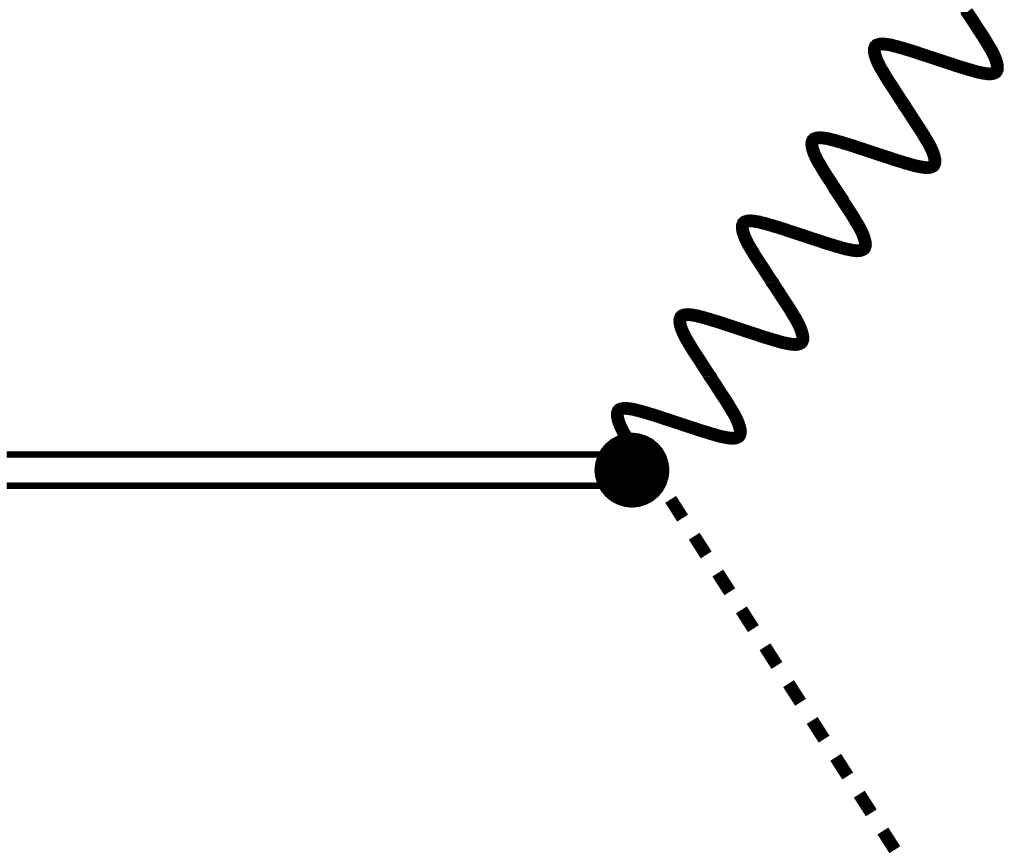}}
\put(-55,20){\makebox(0,0){$\wt a^m$}}
\put(5,-2){\makebox(0,0){$\pi$}}
\put(9,47){\makebox(0,0){$\cV$}}
\put(87,50){\makebox(0,0){$(2)$}}
\put(90,-6){
\includegraphics[scale=0.17]{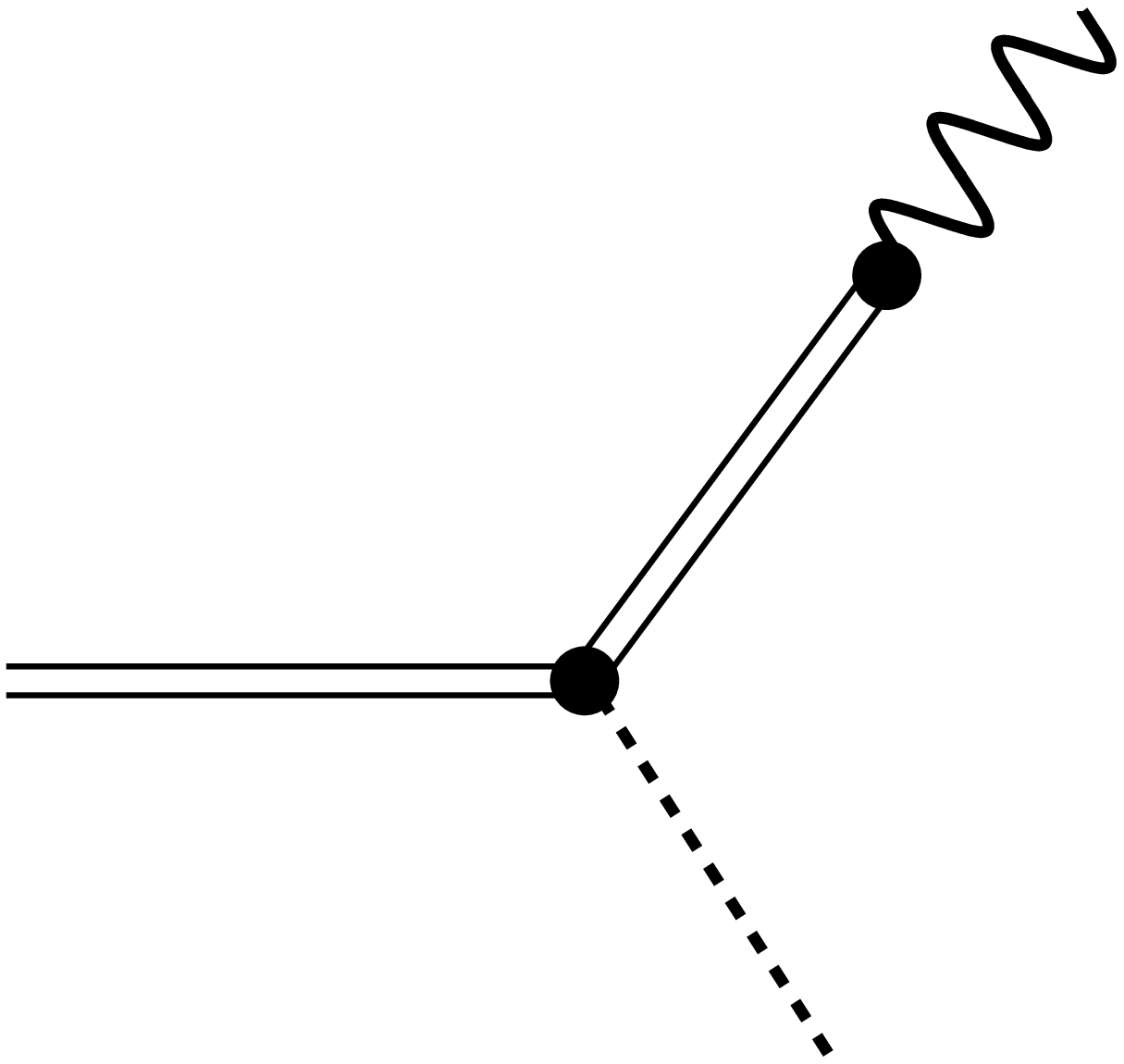}}
\put(85,15){\makebox(0,0){$\wt a^m$}}
\put(145,-8){\makebox(0,0){$\pi$}}
\put(144,22){\makebox(0,0){$\wh v^n$}}
\put(160,54){\makebox(0,0){$\cV$}}
\end{picture}
\caption{\footnotesize{The relevant diagrams for the
$a_1\ra \pi\gamma$ decay amplitude.
}}
\label{fig2}
\end{center}
\end{figure}

First we show that the decay amplitude of
$a_1\ra \pi\gamma$ computed from the effective action in
\S \ref{eff} vanishes.
More generally, we can show that the decay amplitude of
$a^m\ra\pi\gamma$ vanishes for every $m\ge 1$, where $a^m$ is the
axial-vector meson $a_{\mu}^m=B_\mu^{(2m)}$.
The relevant diagrams for this decay amplitude,
depicted in Fig. \ref{fig2}, are
(1) the direct coupling of $\wt a^m\pi\cV$ in (\ref{pi1_2}), which yields
an amplitude proportional to $a_{\cA a^m}$, and
(2) the $\wt a^m\pi \wh v^n$ vertex in (\ref{pi1_2}) accompanied by
the $\wh v^n$-$\cV$ transition in (\ref{L2L3}).
This amplitude is found to be proportional to
\begin{eqnarray}
-\sum_{n\ge 1}c_{v^n a^m\pi}\frac{g_{v^n}}{m_{v^n}^2}=
-\sum_{n\ge 1}c_{v^n a^m\pi} a_{\cV v^n}=
 -a_{\cA a^m} \ ,
\end{eqnarray}
where the sum rule (\ref{aca}) is used.
Therefore, the two diagrams sum to zero.
This fact can be understood more easily from the Lagrangian
(\ref{L2}) and (\ref{pi1}).

The vanishing of the $a_1\ra \pi\gamma$ decay amplitude
has been observed in the HLS model
\cite{BaFuYa,bky;review}
and closely related five-dimensional models \cite{SoSt,DaRoPo,Hirn:2005nr}.
{}From the phenomenological point of view, this is not in
serious conflict
with experiments.
The experimental value of
the partial width of the $a_1\ra\pi\gamma$ decay mode is approximately
640 KeV, while the total width of $a_1$ is $250-600$ MeV \cite{pdg}.
Hence, it seems plausible that the $a_1\ra\pi\gamma$ decay process
is due to the order $1/N_c$ subleading terms or higher derivative terms,
as suggested in Refs.~\citen{BaFuYa} and \citen{bky;review}.

Let us next consider the decay mode $a_1\ra \pi \rho$,
or, more generally, $a^m\ra \pi v^n$.
The decay amplitude can be read from the second
line of (\ref{pi1_2}). Using the equations of motion,
\begin{eqnarray}
\del^\mu(\del_\mu\wh v^n_\nu-\del_\nu\wh v^n_\mu)
&=&m_{v^n}^2\wh v^n_\nu+\cdots\ ,\\
\del^\mu(\del_\mu\wt a^n_\nu-\del_\nu\wt a^n_\mu)
&=&m_{a^n}^2\wt a^n_\nu+\cdots\ ,
\end{eqnarray}
the relevant couplings are extracted as
\begin{eqnarray}
\cL_3|_{\pi^1}\sim\cdots
-2i\,g_{a^m v^n\pi}
\tr\left(
\wt a^m_\mu [\Pi,\wh v^{n\mu}]
\right)+\cdots\ ,
\end{eqnarray}
with
\begin{eqnarray}
g_{a^m v^n\pi}\equiv
\frac{1}{f_\pi}(m_{v^n}^2-m_{a^m}^2)\, c_{v^na^m\pi}\ .
\label{gc}
\end{eqnarray}
Then the decay width is given by \cite{BaFuYa}
\begin{eqnarray}
\Gamma(a^m\ra \pi v^n)=\frac{g_{a^m v^n\pi}^2}{4\pi m_{a^m}^2}
|\bp_{v^n}|
\left(1+\frac{|\bp_{v^n}|^2}{3 m_{v^n}^2}\right)\ ,
\end{eqnarray}
where $\bp_{v^n}$ is the momentum carried by the $v^n$ meson.
{}From the experimental value, \cite{pdg}
\begin{eqnarray}
m_{a^1}|_{\rm exp} \simeq 1230 \mbox{ MeV} \ , ~~
\left.\Gamma(a_1\ra\pi\rho)\right|_{\rm exp}
\simeq 150\sim 360~ \mbox{MeV}\ ,
\end{eqnarray}
the coupling is estimated as
$g_{a_1\rho\pi}^2|_{\rm exp}\simeq 7.6-18~\mbox{GeV}^2$,
and the experimental value of the dimensionless combination
$c_{\rho a_1\pi}$,
which is related to $g_{a_1\rho\pi}$ through (\ref{gc}), is
\begin{eqnarray}
c_{\rho a_1\pi}\equiv
\left. \frac{f_\pi g_{a_1\rho\pi}}{m_{\rho}^2-m_{a_1}^2}
\right|_{\rm exp}\simeq~~ 0.28-0.43\ .
\end{eqnarray}
On the other hand, the numerical analysis of $c_{v^1 a^1\pi}$,
defined in (\ref{cpva}), yields
\begin{eqnarray}
c_{v^1 a^1\pi}\simeq 0.528 \ .
\end{eqnarray}

\subsection{$\pi\pi$ scattering}

It is known that in the chiral limit, the low energy behavior of
the $\pi\pi$ scattering amplitude is governed by only the $\pi^4$
vertex in the lowest derivative term of
the chiral Lagrangian (\ref{Skyrme}).
However, because the $\pi^4$ interaction in (\ref{L4}) is
canceled by that in (\ref{L2L3_2}), one might think that 
the low energy theorem is somehow violated in our model. 
This, of course, is not true.
Here we argue that taking account of the vector meson exchange
diagrams yields
a $\pi\pi$ scattering amplitude that is consistent with
the low energy theorem.

\begin{figure}
\begin{center}
\begin{picture}(100,80)(0,-8)
\put(-110,60){\makebox(0,0){$(1)$}}
\put(-90,20){
\includegraphics[scale=0.19]{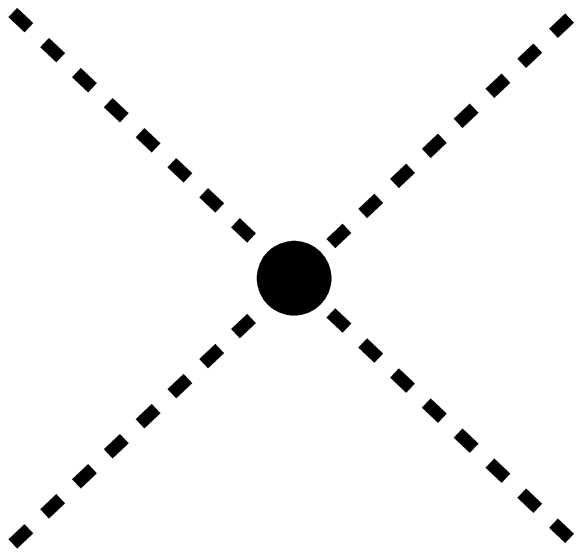}}
\put(-92,50){\makebox(0,0){$\pi$}}
\put(-92,17){\makebox(0,0){$\pi$}}
\put(-50,50){\makebox(0,0){$\pi$}}
\put(-50,17){\makebox(0,0){$\pi$}}
\put(-72,0){\makebox(0,0){$-\frac{1}{3f_\pi^2}$}}
\put(-5,60){\makebox(0,0){$(2)$}}
\put(15,20){
\includegraphics[scale=0.19]{pppp.eps}}
\put(13,50){\makebox(0,0){$\pi$}}
\put(13,17){\makebox(0,0){$\pi$}}
\put(55,50){\makebox(0,0){$\pi$}}
\put(55,17){\makebox(0,0){$\pi$}}
\put(33,0){\makebox(0,0){$+\frac{1}{3f_\pi^2}$}}
\put(103,60){\makebox(0,0){$(3)$}}
\put(118,15){
\includegraphics[scale=0.15]{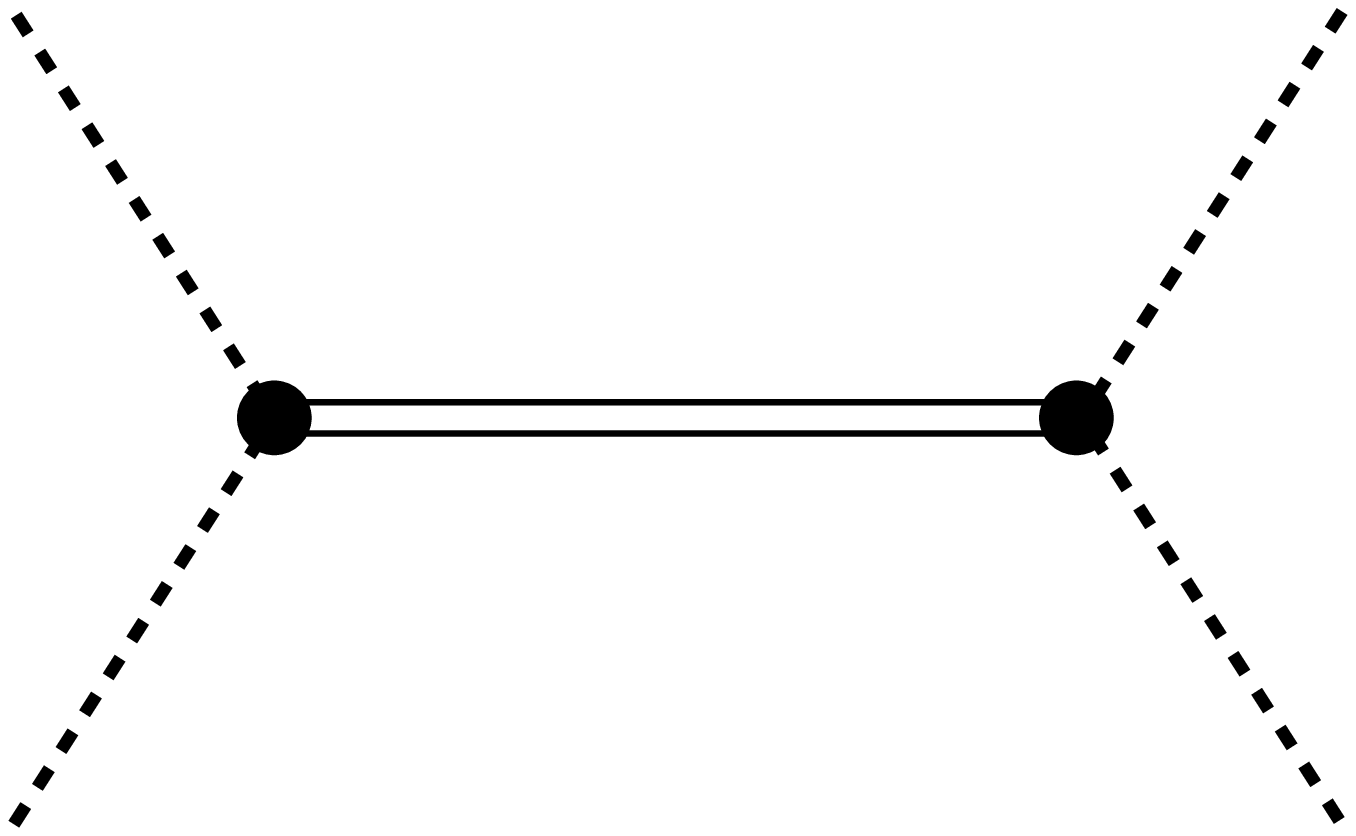}}
\put(118,52){\makebox(0,0){$\pi$}}
\put(118,15){\makebox(0,0){$\pi$}}
\put(185,52){\makebox(0,0){$\pi$}}
\put(185,15){\makebox(0,0){$\pi$}}
\put(155,43){\makebox(0,0){$\wh v^n$}}
\put(150,0){\makebox(0,0)
{$-\frac{g_{v^n\pi\pi}^2}{m_{v^n}^2}$}}
\end{picture}
\caption{\footnotesize{The relevant diagrams for 
the $\pi\pi$ scattering.
}}
\label{fig3}
\end{center}
\end{figure}

The vertices needed to derive the $\pi\pi$ scattering amplitude,
depicted in Fig. \ref{fig3},
consist of (1) the $\pi^4$ couplings in
(\ref{L4}), (2) the direct $\pi^4$ couplings in (\ref{L2L3_2}), and
(3) the $\pi\pi\wh v^n$ couplings in (\ref{L2L3_2}), two of which are
contracted by the vector meson exchanges.
As we have seen in \S \ref{eff}, the contributions from (1) and
(2) cancel. Also, the effective $\pi^4$ vertex obtained
from the exchange of the vector mesons (3) is computed as
\begin{equation}
-\sum_{n=1}^\infty \frac{g_{v^n\pi\pi}^2}{m_{v^n}^2}
\tr [\Pi,\del_\mu\Pi]^2 \ .
\label{pipiv}
\end{equation}
Then, using the sum rule (\ref{sum1}), we end up with 
a term that is identical to the first term (\ref{L4}),
and we thus conclude that the contribution from (3) is
the same as that from (1).
In other words, the contributions from (2) and (3) cancel,
and the low energy $\pi\pi$ scattering amplitude is
governed by the chiral Lagrangian.
This fact is trivial if we use the effective action
given in Appendix \ref{AppA}.
Note that the situation here is very similar
to that in the HLS model with $a=4/3$,
though vector meson dominance does not hold in that case.
(See p.35 of Ref.~\citen{HaYa})

\subsection{Weinberg sum rules}

Before closing this section, let us make a few comments on the
Weinberg sum rules, which turn out to be problematic in our
model. In our notation, the Weinberg sum rules \cite{W:sum} state
\begin{eqnarray}
\sum_{n=1}^\infty
\left(\frac{g_{v^n}^2}{m_{v^n}^2}-\frac{g_{a^n}^2}{m_{a^n}^2}\right)
=f_{\pi}^2 &&~~~[\mbox{Weinberg sum rule (I)}]\ ,
\label{WSM1}\\
\sum_{n=1}^\infty\left(g_{v^n}^2-g_{a^n}^2\right)=0&&
~~~[\mbox{Weinberg sum rule (II)}] \ .
\label{WSM2}
\end{eqnarray}
It was shown in \cite{SoSt} that both (\ref{WSM1}) and (\ref{WSM2})
are satisfied in the discretized version of the five-dimensional
model. On phenomenological grounds, it is often assumed that
the sum rules are almost completely dominated
by the contributions from the $\rho$ and $a_1$ mesons alone, that is,
that we have
\begin{eqnarray}
\frac{g_{\rho}^2}{m_{\rho}^2}-\frac{g_{a_1}^2}{m_{a_1}^2}
\simeq f_{\pi}^2\ ,~~~
g_{\rho}^2\simeq g_{a_1}^2\ .
\label{WSM3}
\end{eqnarray}
In our case, however, the infinite sums in (\ref{WSM1}) and
(\ref{WSM2}) do not converge,
as one can guess from the behavior of the
numerical data (\ref{table}).
Even if this divergence can be removed by appropriately
regularizing the infinite sum, as in Refs.~\citen{SoSt}
and \citen{Hirn:2005nr},
the sums (\ref{WSM1}) and (\ref{WSM2}) are not dominated by
$\rho$ and $a_1$.\footnote{
This problem was pointed out in Ref.~\citen{ChKuTa}
in the context of the discretized model proposed in Ref.~\citen{SoSt}.}
In fact, the ratios of the left-hand sides to the right-hand
sides of the relations given in (\ref{WSM3}) are estimated
in our model as
\begin{eqnarray}
\frac{1}{f_{\pi}^2 }\left(
\frac{g_{v^1}^2}{m_{v^1}^2}-\frac{g_{a^1}^2}{m_{a^1}^2}\right)
\simeq \ -7.38\ ,~~~
\frac{g_{v^1}^2}{g_{a^1}^2} \simeq 0.177\ ,
\end{eqnarray}
which are both far from 1.

Note that the experimental value of
 $g_{a_1}$ estimated using $\tau$ decay \cite{IsMoRe}
is $g_{a_1}|_{\rm exp}\simeq 0.177 \pm 0.014~ \mbox{GeV}^2$,
and the lattice measurement \cite{WiDeCoHe} gives
$g_{a_1}|_{\rm lat}\simeq 0.21 \pm 0.02~\mbox{GeV}^2$.
Both of these values suggest that $g_{a_1}$ is 
larger than $g_{\rho}|_{\rm exp}\simeq 0.12~\mbox{GeV}^2$,
though they are still inconsistent with our numerical result,
 $g_{a_1}/g_{\rho}\simeq 2.38$. It would be interesting
to calculate the corrections in our model
to see if this discrepancy is reconciled.

\section{WZW term}
\label{WZWterm}

In this section, we study the WZW term (\ref{WZW}) to obtain some
interaction terms that involve vector mesons. 
Here again we work in the $A_z=0$ gauge and write the five-dimensional
gauge field in terms of
the differential one-form, $A=A_{\mu}dx^{\mu}+A_zdz=A_{\mu}dx^{\mu}$.
It is useful to first denote the one-form gauge field
in (\ref{5dpot}) or (\ref{exp}) as
\begin{eqnarray}
A&=& v+a\ ,
\label{5dpot:form}
\end{eqnarray}
where
\begin{eqnarray}
v&\equiv&\half(A_++A_-)
+\sum_{n=1}^\infty v^n\,\psi_{2n-1}\ ,
\label{v}\\
a&\equiv&\half(A_+-A_-)\,\psi_0
+\sum_{n=1}^\infty a^n\,\psi_{2n}\ ,
\label{a}
\end{eqnarray}
and
\begin{eqnarray}
A_+\equiv A_{L}^{\xi_+}=
\xi_+A_{L}\xi_+^{-1}+\xi_+ d\xi_+^{-1}\ ,~~~
A_-\equiv A_{R}^{\xi_-}=
\xi_-A_{R}\xi_-^{-1}+\xi_- d\xi_-^{-1}\ .
\end{eqnarray}
Inserting (\ref{5dpot:form}) into
the Chern-Simons 5-form $\omega_5(A)$ in (\ref{WZW}),
we obtain
\begin{eqnarray}
&&\int_{M^4\times\bR}\omega_5(A)\nn\\
&=&
\half
\int_{M^4}\tr\Big[(A_+A_--A_-A_+)\,d(A_++A_-)
+\half A_+A_-A_+A_-+(A_+^3A_--A_-^3A_+)\Big]\nn\\
&&+
\int_{M^4\times\bR} \tr\left(3\,a\,dv\,dv+a\,da\,da+
3\,(v^2 a+a\,v^2+a^3)\,dv+
3\,\Big[a\,v\,a\,da\Big]_{\mbox{\footnotesize non-zero}}\right) \ .
\label{omega}
\end{eqnarray}
(See Appendix \ref{WZW:mani} for details.)
Here, $[\,\cdots\,]_{\mbox{\footnotesize non-zero}}$ denotes 
the contribution from the non-zero modes
that contain terms with at least one vector meson.
It is shown in Ref.~\citen{SaSu} that
the first line in (\ref{omega}), together with the other terms in
(\ref{WZW}), gives the well-known expression of the WZW term
(\ref{WZW0}) that depends only on
the pion field $U$ and the external gauge fields
$A_{L,R}$.
The terms in the second line of (\ref{omega}) are the new terms
that include the interaction with the vector mesons.
As a result, we obtain
\begin{eqnarray}
S_{\rm CS}^{\rm D8}
&=&-\frac{N_c}{48\pi^2}\int_{M^4}Z
-\frac{N_c}{240\pi^2}\int_{M^4\times \bR}\tr(gdg^{-1})^5\nn\\
&&+
\frac{N_c}{24\pi^2}
\int_{M^4\times\bR} \tr\bigg(3\,a\,dv\,dv+a\,da\,da\nn\\
&&~~~~~+3\,(v^2 a+a\,v^2+a^3)\,dv+
3\,\Big[a\,v\,a\,da\Big]_{\mbox{\footnotesize non-zero}}\bigg)\ .
\label{WZWvec}
\end{eqnarray}

In the earlier works concerning the
incorporation of vector mesons into the WZW term,
there are several adjustable parameters that cannot be fixed
by the symmetry in QCD.\cite{FuKuTeUeYa}
 (See also Refs.~\citen{bky;review} and \citen{HaYa}.)
Contrastingly, the couplings including
infinitely many vector mesons in (\ref{WZWvec})
are completely fixed, and there
is no adjustable parameter.
It would therefore be very interesting to determine whether 
a WZW term of the form (\ref{WZWvec}) is consistent
with the experimental data.
In the following subsections, we examine
the phenomenology concerning the $\pi vv$, $\pi v\cV$,
$\pi\cV\cV$, $v\pi^3$ and $\cV\pi^3$ vertices.

\subsection{$\pi vv$, $\pi v\cV$ and $\pi\cV\cV$ vertices}
Note that $\pi vv$ and $\pi v\cV$ couplings appear only
in the first term of the second line in (\ref{omega}).

We set $\cA=0$ for simplicity.
Then, from the expansion (\ref{exp}), we have
\begin{eqnarray}
v&=&\cV+\frac{1}{2f_\pi^2}[\Pi,d\Pi]+\sum_{n=1}^\infty
v^n\psi_{2n-1}+\cdots \ ,\\ 
a&=&\frac{i}{f_\pi}( d\Pi+[\cV,\Pi])\,\psi_0
+\sum_{n=1}^\infty a^n\psi_{2n}+\cdots \ .
\end{eqnarray}
Inserting these forms into (\ref{omega}), we obtain
\begin{eqnarray}
\int_{M^4\times\bR}\omega_5(A)\Big|_{\pi v v,\,\pi v \cV}
&=&\int_{M^4\times\bR}
\tr\left(3\, a\,dv\,dv\right)\Big|_{\pi v v,\,\pi v \cV}\nn\\
&=&-\frac{6i}{f_\pi}\int_{M^4}\tr\left(
\,\Pi\, (dv^n d\cV+d\cV dv^n)\right)\,c_{v^n}\nn\\
&&~~-\frac{6i}{f_\pi}\int_{M^4}\tr\left(
\,\Pi\, dv^n d v^m\right)\, c_{v^n v^m}\ ,
\label{pvv}
\end{eqnarray}
where
\begin{eqnarray}
c_{v^n}&\equiv&
\half\int\! dz\,\del_z{\psi_0}\,\psi_{2n-1}=
\frac{1}{\pi}\int\! dz\,K^{-1}\psi_{2n-1} \ , \\
c_{v^nv^m}&\equiv&\half
\int\! dz\,\del_z{\psi_0}\,\psi_{2n-1}\psi_{2m-1}=
\frac{1}{\pi}\int\!dz\,K^{-1}\psi_{2n-1}\psi_{2m-1}\ .
\end{eqnarray}
Then, using $\wt v^n$ defined in (\ref{wtv}) and the sum rules
\begin{eqnarray}
\sum_{m=1}^\infty a_{\cV v^m}c_{v^n v^m}=c_{v^n}\ ,~~~
\sum_{n=1}^\infty a_{\cV v^n}c_{v^n}=1\ ,
\label{acc}
\end{eqnarray}
(\ref{pvv}) can be written
\begin{eqnarray}
\int_{M^4\times\bR}\omega_5(A)\Big|_{\pi v v,\,\pi v \cV}
=\frac{6i}{f_\pi}\int_{M^4}\tr\left(
\,\Pi\, d\cV d\cV\, \right)
-\frac{6i}{f_\pi}\int_{M^4}\tr\left(
\,\Pi\, d\wt v^n d\wt v^m\right)\, c_{v^n v^m}\ .
\label{pvv2}
\end{eqnarray}

It is easy to check that the contributions to
the $\pi^0\ra \gamma\gamma$ decay amplitude from the two terms
in (\ref{pvv2}) cancel.
This is obvious from the fact that there is no $\pi\cV\cV$
vertex in (\ref{pvv}) that is written
in terms of $v^n$ and $\cV$.

The $\pi\cV\cV$ vertex comes from $Z$ given in (\ref{Z}):
\begin{eqnarray}
\int_{M^4} Z\,\Big|_{\pi \cV\cV}
&=&\int_{M^4}\Big(\tr\left[(A_R dA_R+dA_R A_R)U^{-1}dU
-\mbox{p.c.}\right]\nn\\
&&~~+\tr\left[ dA_R dU^{-1}A_L U-\mbox{p.c.}\right]
\Big)\Big|_{\pi \cV\cV}\nn\\
&=&\frac{12i}{f_\pi}\int_{M^4}\tr\left(\,\Pi\, d\cV d\cV\,\right) \ .
\label{pvvZ}
\end{eqnarray}
Combining (\ref{pvv2}) and (\ref{pvvZ}), and, furthermore,
rewriting them in terms of $\wt v^n$ given in (\ref{wtv}) or
$\wh v^n$ given in (\ref{whv}),
we obtain
\begin{eqnarray}
S_{\rm CS}^{\rm D8}\Big|_{\pi vv,\,\pi v\cV,\,\pi\cV\cV}
&=&-\frac{N_c}{48\pi^2}\int_{M^4}Z\,\Big|_{\pi\cV\cV}
+\frac{N_c}{24\pi^2}\int\omega_5(A^g)
\Big|_{\pi v v,\,\pi v \cV}\nn\\
&=&-\frac{N_c}{4\pi^2}\frac{i}{f_\pi}
\int_{M^4}\tr\left(\Pi\, d\wt v^n d\wt v^m\right)\,c_{v^n v^m}\nn\\
&=&-\frac{N_c}{4\pi^2}\frac{i}{f_\pi}
\int_{M^4}\tr\left(\Pi\, d\wh v^n d\wh v^m\right)\,c_{v^n v^m}\nn\\
&&~~-\frac{N_c}{2\pi^2}\frac{i}{f_{\pi}^3}(c_n-d_n)
\int_{M^4}\tr\left( d\Pi\, d\Pi\, d\Pi\, \wh v^n\right)
+\cO(\Pi^5)\ ,
\label{WZW:pivv}
\end{eqnarray}
where
\begin{eqnarray}
d_{v^n}\equiv\frac{1}{2}
\int\!dz\,\psi_0^2\del_z{\psi_0}\,\psi_{2n-1}=
\frac{1}{\pi}\int\!dz\,K^{-1}\psi_0^2\psi_{2n-1}\ ,
\end{eqnarray}
and we have used the sum rule
\begin{equation}
\sum_{m=1}^\infty
b_{v^m\pi\pi}c_{v^nv^m}=c_{v^n}-d_{v^n} \ .
\label{bccd}
\end{equation}
We thus conclude that there exist no direct three-point
couplings including the external photon field. This demonstrates
the vector meson dominance in this sector.

\subsection{$v\pi^3$ and $\cV\pi^3$ vertices}
\label{vppp}
The $v^n\pi^3$ vertex comes from the $3\,a\,dv\,dv$, $3\,a^3dv$
and $3[a\,v\,a\,da]_{\mbox{\footnotesize non-zero}}$
terms in (\ref{omega}):
\begin{eqnarray}
\int_{M^4\times\bR}\tr(3\,a\,dv\,dv)\Big|_{v\pi^3}
=\frac{12i}{f_\pi^3}\int_{M^4}\tr\left(
\,d\Pi\, d\Pi\, d\Pi\, v^n\right)c_{v^n} \ ,
\label{pppv1}
\end{eqnarray}
\begin{eqnarray}
\int_{M^4\times\bR}
\tr\left(3\,a^3dv+3\Big[a\,v\,a\,da\Big]_{\mbox{\footnotesize non-zero}}
\right)\Bigg|_{v\pi^3}
=-\frac{12i}{f_\pi^3}\int_{M^4}
\tr\left(d\Pi\,d\Pi\,d\Pi\,v^n\right)\,d_{v^n}\ .
\label{pppv2}
\end{eqnarray}
{}From (\ref{pppv1}) and  (\ref{pppv2}), we obtain
\begin{eqnarray}
\int_{M^4\times\bR}\omega_5(A)\Big|_{v \pi^3}
=\frac{12i}{f_\pi^3}(c_{v^n}-d_{v^n})
\int_{M^4}\tr(d\Pi\,d\Pi\,d\Pi\,v^n) \ .
\label{pppv3}
\end{eqnarray}
Rewriting this relation in terms of $\wt v^n$ in (\ref{wtv}),
and using the sum rules
\begin{eqnarray}
\sum_{n=1}^\infty a_{\cV v^n}c_{v^n}=1\ ,~~~
\sum_{n=1}^\infty a_{\cV v^n}d_{v^n}
=\frac{1}{\pi}\int\!dz\,K^{-1}\psi_0^2
=\frac{1}{3}\ ,
\end{eqnarray}
we obtain
\begin{eqnarray}
\int_{M^4\times\bR}\omega_5(A)\Big|_{v \pi^3}
=\frac{12i}{f_\pi^3}(c_{v^n}-d_{v^n})
\int_{M^4}\tr(d\Pi\,d\Pi\,d\Pi\,\wt v^n)
-\frac{8i}{f_\pi^3}
\int_{M^4}\tr(d\Pi\,d\Pi\,d\Pi\,\cV) \ .\nn\\
\label{pppv4}
\end{eqnarray}
The $\cV\pi^3$ vertex can be read off of $Z$ in (\ref{Z}):
\begin{eqnarray}
\int_{M^4}Z\,\Big|_{\cV\pi^3}&=&
\int_{M^4}\tr\left[A_R(dU^{-1}U)^3-{\rm p.c.}\right]
\Big|_{\cV\pi^3}\nn\\
&=&-\frac{16i}{f_\pi^3}\int_{M^4}\tr\left(
d\Pi\,d\Pi\,d\Pi\,\cV\right) \ .
\label{pppvZ}
\end{eqnarray}
Collecting (\ref{pppv4}) and (\ref{pppvZ}), we obtain
\begin{eqnarray}
S_{\rm CS}^{\rm D8}
\Big|_{v\pi^3,\,\cV\pi^3}
&=&-\frac{N_c}{48\pi^2}\int_{M^4}Z\,
\Big|_{\cV\pi^3}
+\frac{N_c}{24\pi^2}\int_{M^4\times\bR}\omega_5(A)
\Big|_{v\pi^3,\,\cV\pi^3}\nn\\
&=&\frac{N_c}{2\pi^2}\frac{i}{f_\pi^3}(c_{v^n}-d_{v^n})
\int_{M^4}\tr(d\Pi\,d\Pi\,d\Pi\,\wt v^n) \ .
\label{pppwtv}
\end{eqnarray}
This again exhibits the vector meson dominance.
Moreover, if we write the action in terms of
$\wh v^n$, the $\wh v^n \pi^3$ coupling in (\ref{pppwtv})
cancels that in (\ref{WZW:pivv}), and we finally obtain
\begin{eqnarray}
S_{\rm CS}^{\rm D8}\Big|_{\pi vv,\,\pi v\cV,\,\pi\cV\cV,
v\pi^3,\cV\pi^3}
&=&
 -\frac{N_c}{4\pi^2}\frac{i}{f_\pi}
\int_{M^4}
\tr\left(\Pi\, d\wh v^n d\wh v^m\right) c_{v^nv^m}
+\cO(\Pi^5)\ .
\label{pwhvwhv}
\end{eqnarray}

\subsection{$\omega\ra\pi^0\gamma$
 and $\omega\ra\pi^0\pi^+\pi^-$ decay}

\begin{figure}
\begin{center}
\begin{picture}(100,80)(0,-10)
\put(-77,50){\makebox(0,0){$(1)$}}
\put(-60,-6){
\includegraphics[scale=0.17]{a_pi_v_gamma.eps}}
\put(-65,15){\makebox(0,0){$\omega$}}
\put(-5,-8){\makebox(0,0){$\pi$}}
\put(-8,20){\makebox(0,0){$\wh v^n$}}
\put(8,54){\makebox(0,0){$\gamma$}}
\put(65,50){\makebox(0,0){$(2)$}}
\put(70,-6){
\includegraphics[scale=0.17]{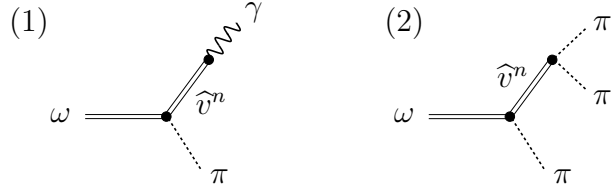}}
\put(65,15){\makebox(0,0){$\omega$}}
\put(106,30){\makebox(0,0){$\wh v^n$}}
\put(125,-8){\makebox(0,0){$\pi$}}
\put(140,22){\makebox(0,0){$\pi$}}
\put(140,50){\makebox(0,0){$\pi$}}
\end{picture}
\caption{\footnotesize{The relevant diagrams for 
(1) $\omega\ra\pi\gamma$
and (2) $\omega\ra\pi\pi\pi$.
}}
\label{fig4}
\end{center}
\end{figure}

{}From the coupling (\ref{pwhvwhv}), we can calculate
the $\omega\ra \pi^0\gamma$ and $\omega\ra\pi^0\pi^+\pi^-$
decay amplitudes.
Here, $\omega$ is the iso-singlet component of the lightest vector meson,
$\wh v^1$.
Because of the complete vector meson dominance
and the absence of the direct $\wh v^n\pi^3$ coupling,
the former is given by the vertex $\omega\ra\wh v^n\rho$,
followed by the $\wh v^n\ra\gamma$ transition, and the latter is given by 
$\omega\ra \pi\wh v^n$, followed by $\wh v^n\ra2\pi$
 (see Fig. \ref{fig4}).
These diagrams are identical to those in the Gell-Mann -
Sharp - Wagner (GSW) model \cite{GSW}, which is
known to be in good agreement with experimental data.
Let us examine how it works in our model. 

The calculation of the $\omega\ra \pi^0\gamma$ decay amplitude
is analogous to that given in Refs.~\citen{FuKuTeUeYa} and
\citen{HaYa}, and we obtain
\begin{eqnarray}
\Gamma(\omega\ra\pi^0\gamma)= 
\frac{N_c^2}{3}\frac{\alpha}{64\pi^4 f_\pi^2}
\left(
\sum_{m=1}^\infty\frac{c_{v^1v^m}g_{v^m}}{m_{v^m}^2}
\right)^2 |\bp_\pi|^3
=
\frac{N_c^2}{3}\frac{\alpha}{64\pi^4 f_\pi^2}
c_{v^1}^2 |\bp_\pi|^3
\ ,
\end{eqnarray}
where $\alpha=e^2/4\pi$, and
we have used the sum rule (\ref{acc}).
Here, $c_{v^1}$ plays the role of the parameter $g$
in Ref.~\citen{FuKuTeUeYa} and it is shown there
that the decay width is consistent with the experimental value when
$g\simeq g_{\rho\pi\pi}$.\footnote{
A recent experimental value \cite{pdg} is 
$\left.\Gamma(\omega \ra \pi^0\gamma)\right|_{\rm exp}
\simeq0.757~\mbox{MeV}$, which implies
$c_{\omega}|_{\rm exp}\simeq 5.80$.}
Remarkably, this is exactly what we have in our model.
In fact, it can easily be shown that, in general,
\begin{eqnarray}
c_{v^n}=g_{v^n\pi\pi}\ ,
\label{cg}
\end{eqnarray}
using (\ref{deqn;psi}) and integrating by parts in the expression
for $g_{v^n\pi\pi}$ given by (\ref{couplings}) and (\ref{vnpp}).

Similarly, the $\omega\ra\pi^0\pi^+\pi^-$ decay width is
\cite{FuKuTeUeYa,HaYa}
\begin{eqnarray}
\Gamma(\omega\ra\pi^0\pi^+\pi^-)=\frac{m_\omega}{192\pi^3}
\int\!\!\int\!
 dE_+dE_-\left[|{\bq_-}|^2|{\bq_+}|^2-({\bq_+}\cdot{\bq_-})^2
 \right]|F_{\omega\ra 3\pi}|^2\ ,
\label{oppp}
\end{eqnarray}
where $E_\pm$ are the energies  and ${\bq}_\pm$
are the momenta of $\pi^\pm$ in the rest frame of $\omega$,
and
\begin{eqnarray}
F_{\omega\ra 3\pi}&\equiv&-\frac{N_c}{4\pi^2f_\pi}
\sum_{n=1}^\infty c_{v^1v^n} g_{v^n\pi\pi}\nn\\
&&~~~\times\left(\frac{1}{m_{v^n}^2+(q_++q_-)^2}
+\frac{1}{m_{v^n}^2+(q_-+q_0)^2}+\frac{1}{m_{v^n}^2+(q_0+q_+)^2}
\right)\ ,
\label{Foppp}
\end{eqnarray}
where $q_0$ and $q_\pm$ are the four-momenta of $\pi^0$
and $\pi^\pm$, respectively.
The results of our numerical analysis suggest that the $n=1$ term
dominates the sum.
If we replace the entire sum with this single term,
the expression for the decay width in (\ref{oppp}) becomes
the same as that in Ref.~\citen{HaYa} with the parameters $g$
and $c_i$ ($i=1,2,3$)
chosen as
\begin{eqnarray}
g=\frac{2f_\pi^2 g_{v^1\pi\pi}}{m_{v^1}^2}c_{v^1v^1}
=b_{v^1\pi\pi}c_{v^1v^1}\ ,
\label{gcb}
\end{eqnarray}
and $c_1-c_2=c_3=1$. It is shown in Refs.~\citen{FuKuTeUeYa}
and \citen{HaYa} that
the decay width is consistent with the experimental results
again with $g\simeq g_{\rho\pi\pi}$.
Note that the right-hand side of the relation (\ref{gcb})
is the main contribution on the left-hand side of (\ref{bccd}).
If we approximate (\ref{gcb}) with (\ref{bccd}),
the relation (\ref{gcb}) is replaced by
\begin{eqnarray}
g\simeq c_{v^1}-d_{v^1}=g_{v^1\pi\pi}-d_{v^1}\ ,
\label{gcb2}
\end{eqnarray}
where we have used (\ref{cg}).
We have found numerically that $d_{v^1}$ is much smaller than $c_{v^1}$,
as seen in (\ref{table2}). This implies that $g\simeq g_{\rho\pi\pi}$,
as desired.

Here we list the results of our numerical estimations
of several quantities:
\begin{eqnarray}
\begin{array}{c||cccccc}
n&\kappa\,\, c_{v^1v^n}&\kappa^{-1/2}b_{v^n\pi\pi}
&\kappa^{1/2} c_{v^n}&\kappa^{1/2} d_{v^n}\\
\hline
1&0.202&1.58&0.415&0.0875
\\
2&-0.0992&-0.0964&-0.109&0.0485\\
3&0.0284&0.00619&0.0160&-0.0367\\
4&-0.00618&-0.000880&-0.00408&0.00887
\end{array}
\label{table2}
\end{eqnarray}
As a check, we note that the numerically determined value of
(\ref{gcb}) is $g\simeq 0.319\,\kappa^{-1/2}$, while that of
(\ref{gcb2}) is $g\simeq 0.328\,\kappa^{-1/2}$.
It is found that the above approximation gives reasonable results,
though they are not extremely close to
$g_{v^1\pi\pi}\simeq 0.415\,\kappa^{-1/2}$.
If we use the value of $\kappa$ in (\ref{kappa}),
the parameter $g$ in (\ref{gcb}) is estimated as $g\simeq 3.69$,
which is about 64\% of the experimental value of
$g_{\rho\pi\pi}$.

To determine the extent to which the result is affected by including
the contributions from $n> 1$ terms in (\ref{Foppp}),
let us estimate the decay width by performing
the integration in (\ref{oppp}).
By using (\ref{kappa}) and the experimental values
$N_c=3$, $m_{\pi^{\pm}}\simeq 140~\mbox{MeV}$ and
$m_{\pi^0}\simeq 135~\mbox{MeV}$, we obtain
\begin{eqnarray}
\Gamma_k(\omega\ra\pi^+\pi^-\pi^0)\simeq
\{2.48,~2.58,~2.58,~2.58,\cdots\}~\mbox{MeV}\ .
\label{Gk}
\end{eqnarray}
Here, $\Gamma_k(\omega\ra 3\pi)$ denotes the decay width with the exchange
of the $v^n$ $(1\le n\le k)$ vector mesons incorporated in (\ref{Foppp}).
Therefore, the contribution of the $\rho$ meson exchange dominates
the sum. Unfortunately, this value is much smaller than
the experimental value, $\Gamma(\omega\ra\pi^+\pi^-\pi^0)|_{\rm exp}
\simeq 7.56~\mbox{MeV}$.\cite{pdg} This is mainly due to the
smallness of the coupling $g$ estimated above.

\section{Vector meson dominance revisited}
\label{revisit}
In the previous sections, we have shown
that our model exhibits the vector meson dominance
by examining the couplings with the external gauge fields one by one.
Here we present a more systematic way to understand
why it works.

In this subsection, we work with 
the expansions given in (\ref{extexp}) and (\ref{exp2}).
We can gauge away $\varphi^{(n)}$ in (\ref{exp2})
without changing the asymptotic condition (\ref{asymp}).
Then, these expansions are written as
\begin{eqnarray}
A_\mu&=&\cV_\mu+\cA_\mu\psi_0
+\sum_{n=1}^\infty v^n_\mu\,\psi_{2n-1}
+\sum_{n=1}^\infty a^n_\mu\,\psi_{2n}\ ,
\label{extexp2}\\
A_z&=&-i\,\Pi\, \phi_0\ ,
\label{AzPi}
\end{eqnarray}
where $\Pi$ denotes the pion field $\varphi^{(0)}$ in (\ref{exp2}).
Note that $v^n_\mu$, $a^n_\mu$ and $\Pi$ in these expansions
are not exactly equal to those appearing in \S \ref{DBIpart}
and \S \ref{WZWterm},
but they are related through certain field redefinitions.
If we rewrite the expansion (\ref{extexp2})
in terms of $\wt v^n_\mu$ and $\wt a^n_\mu$
given in (\ref{wtv}) and (\ref{wta}), we have
\begin{eqnarray}
A_\mu
&=&\cV_\mu\,\psi_v+\cA_\mu\psi_a
+\sum_{n=1}^\infty\wt v^n_\mu\,\psi_{2n-1}
+\sum_{n=1}^\infty\wt a^n_\mu\,\psi_{2n}\ ,
\label{extexp3}
\end{eqnarray}
where\footnote{The infinite sums in (\ref{psi_va}) should be regarded
as formal expressions, since they do not uniformly converge to smooth,
normalizable functions.
In the following, we take the limit of
an infinite sum after performing the integration over $z$.}
\begin{eqnarray}
\psi_v\equiv1-\sum_{n=1}^\infty a_{\cV v^n}\psi_{2n-1}\ ,~~~
\psi_a\equiv\psi_0-\sum_{n=1}^\infty a_{\cA a^n}\psi_{2n}\ .
\label{psi_va}
\end{eqnarray}
Note that it can be shown using (\ref{norm;psi}) 
and (\ref{aVvaAa}) that
\begin{eqnarray}
0=\int\! dz\, K^{-1/3}\psi_v\psi_{m}
=\int\! dz\, K^{-1/3}\psi_a\psi_{m}\ 
\end{eqnarray}
for all $m$. Equivalently, we have
\begin{eqnarray}
0=\int\! dz\, K^{-1/3}\psi_v f=\int\! dz\, K^{-1/3}\psi_a f
\label{ortho}
\end{eqnarray}
for an arbitrary normalizable function $f(z)$.
{}From this fact, we immediately see
that if we write the action in terms
of $\wt v^n_\mu$ and $\wt a^n_\mu$
defined in (\ref{wtv}) and (\ref{wta}),
many of the couplings that include
$\cV_\mu$ or $\cA_\mu$ vanish.
In fact, following the procedure described in Appendix \ref{AppA},
we obtain
\begin{eqnarray}
&&\kappa\int\!dz\,\tr\left[\half K^{-1/3} F_{\mu\nu}^2\right]
\nn\\
&=&\tr\Bigg[\,
\frac{1}{2e^2}\left(
(F_{\mu\nu}^{A_L})^2+(F_{\mu\nu}^{A_R})^2\right)\nn\\
&&
+\half(\del_\mu \wt v_\nu^n-\del_\nu \wt v_\mu^n)^2
+\half(\del_\mu \wt a_\nu^n-\del_\nu \wt a_\mu^n)^2\nn\\
&&
+(\del_\mu \wt v_\nu^n-\del_\nu \wt v_\mu^n)
([\wt v^{p\mu},\wt v^{q\nu}]\, g_{v^nv^pv^q}+
[\wt a^{p\mu},\wt a^{q\nu}]\, g_{v^na^pa^q})\nn\\
&&
+(\del_\mu \wt a_\nu^n-\del_\nu \wt a_\mu^n)
([\wt v^{p\mu},\wt a^{q\nu}] -
[\wt v^{q\nu},\wt a^{p\mu}] )\,g_{v^pa^na^q}\nn\\
&&+
\half[\wt v^m_\mu,\wt v^n_\nu][\wt v^{p\mu},\wt v^{q\nu}]
\,g_{v^mv^nv^pv^q}
+\half[\wt a^m_\mu,\wt a^n_\nu][\wt a^{p\mu},\wt a^{q\nu}]
\,g_{a^ma^na^pa^q}\nn\\
&&
+\big([\wt v^m_\mu,\wt v^n_\nu][\wt a^{p\mu},\wt a^{q\nu}]+
[\wt v^m_\mu,\wt a^p_\nu][\wt v^{n\mu},\wt a^{q\nu}]
-[\wt v^m_\mu,\wt a^p_\nu][\wt v^{n\nu},\wt a^{q\mu}]
\big)
\,g_{v^mv^na^pa^q}
\Bigg]\ .
\end{eqnarray}
[ See (\ref{def:g}) for the definition of the four-point coupling 
constants. ]
This shows that all the couplings between the external
gauge fields $(A_L,A_R)$ and the vector meson fields
$(\wt v^n,\wt a^n)$ vanish
in the first term of the effective action (\ref{DBI}).
Here we have used the relations
\begin{eqnarray}
\kappa\int\!dz\,K^{-1/3}\psi_a^{2n}\psi_v^m&=&
\kappa\int\!dz\,K^{-1/3}\psi_a\psi_0\ ,
~~~~(\mbox{for $n\ge 1$, $m\ge 0$})\\
\kappa\int\!dz\,K^{-1/3}\psi_a^{2n}\psi_v^m&=&
\kappa\int\!dz\,K^{-1/3}\psi_v\ ,
~~~~(\mbox{for $m\ge 1$, $n\ge 0$})
\end{eqnarray}
which can be formally shown by using (\ref{ortho}), and we have set
\begin{eqnarray}
e^{-2}\equiv\frac{\kappa}{2}\int\! dz\, K^{-1/3}
\psi_v\nn\ .
\end{eqnarray}
This is divergent (or ill-defined), because
$\psi_v$ is a function that approaches 1 at $z\ra\pm\infty$.

It is important to note that we cannot conclude that $\psi_v=\psi_a=0$
{}from the relation (\ref{ortho}). Actually, using (\ref{deqn;psi})
and (\ref{norm;psi}), one can show
\begin{eqnarray}
\kappa\int\!dz\, K\,\del_z\psi_v\,\del_z\psi_{2n-1}&=&
-\lambda_{2n-1}a_{\cV v^n}\ ,\\
\kappa\int\!dz\, K\,\del_z\psi_a\,\del_z\psi_{2n}&=&
-\lambda_{2n}a_{\cA a^n}\ .
\end{eqnarray}
Then, the second term in the effective action (\ref{DBI}) is 
calculated as
\begin{eqnarray}
\kappa\int\!dz\,\tr\left[ K F_{z\nu}^2\right]
&=&\tr\bigg[\,m_{v^n}^2(\wt v^n_\mu-a_{\cV v^n}\cV_\mu)^2+
m_{a^n}^2(\wt a^n_\mu-a_{\cA a^n}\cA_\mu)^2+
(i\del_\mu\Pi+f_\pi\cA_\mu)^2\nn\\
&&~~~+2i g_{a^mv^n\pi}\,\wt a^m_\mu[\Pi,\wt v^{n\mu}]
-2g_{v^n\pi\pi}\,\wt v^n_\mu[\Pi,\del^\mu\Pi]
\nn\\
&&~~~-c_{a^na^m}[\Pi,\wt a^n_\mu][\Pi,\wt a^{m\mu}]-
c_{v^nv^m}[\Pi,\wt v^n_\mu][\Pi,\wt v^{m\mu}]
\,\bigg]\ ,
\label{Lsec}
\end{eqnarray}
where
\begin{eqnarray}
c_{a^na^m}\equiv\frac{1}{\pi}\int\!dz\, K^{-1}\psi_{2n}\psi_{2m}\ .
\end{eqnarray}
Here, we have used the relation (\ref{cg}), as well as
the fact that $g_{a^mv^n\pi}$
defined in (\ref{gc}) is equal to
\begin{eqnarray}
g_{a^mv^n\pi}=f_\pi
\int\!dz\,\psi_{2m}\del_z\psi_{2n-1}\ ,
\end{eqnarray}
which can be shown by using (\ref{deqn;psi}).
The mesons couple to the external gauge fields only through
the $\wt v^n\ra\cV$ and $\wt a^n\ra\cA$ transitions in
(\ref{Lsec}).

The expression for the WZW term can also be simplified
by using the expansions (\ref{extexp3}) and (\ref{AzPi}).
It is easy to see that
all the terms including $\cV_\mu$ and $\cA_\mu$ vanish because
of the relation (\ref{ortho}). This demonstrates the complete
vector meson dominance in the WZW term.
Moreover, because the pion field $\Pi$ appears only in (\ref{AzPi}),
the terms with two or more pion fields vanish, as we
partly observed in \S \ref{vppp}.
Inserting  (\ref{extexp3}) and (\ref{AzPi}) into (\ref{CS}),
we obtain
\begin{eqnarray}
S_{\rm D8}^{\rm CS}
&=&
-\frac{N_c}{4\pi^2}\frac{i}{f_\pi}\int_{M^4} \tr\Big[\,
\Pi\, dB^ndB^m\, c_{nm}\nn\\
&&~~~~~~+\Pi\,(dB^mB^nB^p+B^mB^ndB^p)\,c_{mnp}
+\Pi\,B^mB^nB^pB^q\,c_{mnpq}\Big]\nn\\
&&~~~+
\frac{N_c}{24\pi^2}
\int_{M^4} \tr\Big[\,B^mB^ndB^p\,d_{mn|p}
-\frac{3}{2}B^mB^nB^pB^q\, d_{mnp|q}\Big]\ ,
\end{eqnarray}
where $B^{2n-1}\equiv\wt v^n$, $B^{2n}\equiv\wt a^n$
and
\begin{eqnarray}
&&c_{mn}\equiv\frac{1}{\pi}\int\!dz\,K^{-1}\psi_m\psi_n\ , ~~
c_{mnp}\equiv\frac{1}{\pi}\int\!dz\,K^{-1}\psi_m\psi_n\psi_p\ , \nn\\
&&c_{mnpq}\equiv\frac{1}{\pi}\int\!dz\,K^{-1}\psi_m\psi_n\psi_p\psi_q\ ,
\nn\\
&&d_{mn|p}\equiv\int\!dz\,(\psi_n\del_z\psi_m-\psi_m\del_z\psi_n)\,
\psi_p\ ,
~~~
d_{mnp|q}\equiv\int\!dz\,\psi_m\psi_n\psi_p\,\del_z\psi_q\ .
\end{eqnarray}

\section{Summary and discussion}
\label{discussion}
In this paper, we computed the effective action 
including the pion, the vector mesons and the external gauge fields
associated with the chiral $U(N_f)_L\times U(N_f)_R$ symmetry,
based on the D4/D8 model proposed in Ref.~\citen{SaSu}, which is
conjectured to be a holographic dual of large $N_c$ QCD
with $N_f$ massless flavors.
We estimated various coupling constants numerically and
compared these values with the experimental values.
The agreement was, of course, not perfect, but we think that it is
good enough to believe that our model nicely captures the expected
features of QCD even quantitatively.

One of the major issues addressed in this paper is vector
meson dominance. An intuitive explanation for this phenomenon
in our model is as follows (see also Ref.~\citen{DaRoPo}).
As seen in \S \ref{revisit},
the external gauge fields appearing in (\ref{extexp3}) have
support only at the boundary, corresponding to $z\ra\pm\infty$,
while the pion and the vector mesons
correspond to the normalizable modes, which vanish as $z\ra\pm\infty$.
Therefore the external gauge fields cannot couple to the mesons,
unless the divergent factor $K$ in the second term of
(\ref{DBI}) picks up the contribution in the $z\ra\pm\infty$ limit.
For the second term of (\ref{DBI}), we know that the fields
$v^n_\mu$ are the mass eigenmodes and hence that
the fields $\wt v^n_\mu$ mix with $\cV_\mu$.

We have found various useful sum rules among the masses and
the coupling constants of an infinite tower of vector
mesons. These follow from the completeness condition of
the mode functions (\ref{comp}). As far as we have determined,
the contribution from the $\rho$ meson is always
the most dominant term in the sum rules. Approximating
these infinite sums with the contribution from only the $\rho$ meson,
we obtained KSRF-type relations in \S \ref{KSRF} and, furthermore,
the approximate $\rho$ meson dominance and $\rho$
meson coupling universality, as explained
in \S \ref{s:VMD}.

In order to make more reliable predictions,
we must take into account the string loop corrections
and the $\alpha'$ corrections  and also
go beyond the probe approximation.
It would be quite interesting to see how our results for
the KSRF relations, the Weinberg sum rules, the $a_1$ meson decay
amplitudes, etc., are improved by incorporating such corrections.
(See a forthcoming paper, Ref.~\citen{HaMaYa} for a discussion along
this line.)

\section*{Acknowledgements}

We would like to thank our colleagues at the Yukawa Institute for
Theoretical Physics and the particle theory group
at Ibaraki University for discussions and encouragement. 
S. S. is especially grateful to 
M. Harada, M. Kurachi, S. Matsuzaki, T. Onogi,
D.T. Son, M. Tachibana, M. Tanabashi and K. Yamawaki
for various useful and enjoyable discussions.
T. S. would like to thank T. Fujiwara for discussions.
The work of S. S. is supported by a Grant-in-Aid for the 21st Century COE
``Center for Diversity and Universality in Physics'' from the Ministry
of Education, Culture, Sports, Science and Technology (MEXT) of Japan.

\appendix

\section{The effective action (DBI part)}
\label{AppA}
Here we calculate the field strength of the gauge field
appearing in (\ref{5dpot}) and (\ref{exp}):
\begin{eqnarray}
A_\mu
&=&\half(A_{L\mu}^{\xi_+}+A_{R\mu}^{\xi_-})
+\half(A_{L\mu}^{\xi_+}-A_{R\mu}^{\xi_-})\psi_0
+\sum_{n=1}^\infty v^n_\mu\,\psi_{2n-1}
+\sum_{n=1}^\infty a^n_\mu\,\psi_{2n}\nn\\
&=&\wh\cV_\mu+\wh\cA_\mu\psi_0
+\sum_{n=1}^\infty v^n_\mu\,\psi_{2n-1}
+\sum_{n=1}^\infty a^n_\mu\,\psi_{2n} \ ,
\end{eqnarray}
where
\begin{eqnarray}
\wh\cV_\mu\equiv\half(A_{L\mu}^{\xi_+}+A_{R\mu}^{\xi_-})\ ,~~~
\wh\cA_\mu\equiv\half(A_{L\mu}^{\xi_+}-A_{R\mu}^{\xi_-})\ .
\label{whcVwhcA}
\end{eqnarray}
In the $\xi_+^{-1}=\xi_-=e^{i\Pi/f_\pi}$ gauge, we can expand
these fields as
\begin{eqnarray}
\wh\cV_\mu&=&
\cV_\mu+\frac{1}{2f_\pi^2}[\,\Pi,\del_\mu\Pi\,]
-\frac{i}{f_\pi}[\,\Pi,\cA_\mu\,]+\cdots \ ,\\
\wh\cA_\mu&=&
\cA_\mu+\frac{i}{f_\pi}\del_\mu\Pi
-\frac{i}{f_\pi}[\,\Pi,\cV_\mu\,]+\cdots \ .
\end{eqnarray}
Then, the field strengths are obtained as
\begin{eqnarray}
F_{z\mu}=\del_z A_\mu=
\wh\cA_\mu\frac{2}{\pi K}
+\sum_{n=1}^\infty v^n_\mu\,\del_z\psi_{2n-1}
+\sum_{n=1}^\infty a^n_\mu\,\del_z\psi_{2n} \ ,
\end{eqnarray}
and
\begin{eqnarray}
F_{\mu\nu}=\del_\mu A_\nu-\del_\nu A_\mu+[A_\mu,A_\nu]
&=&F_{\mu\nu}\Big|_{\rm even}+F_{\mu\nu}\Big|_{\rm odd}\ ,
\end{eqnarray}
\begin{eqnarray}
F_{\mu\nu}\Big|_{\rm even}&=&
F_{\mu\nu}^{\wh\cV}+[\wh\cA_\mu,\wh\cA_\nu]\psi_0^2+
\sum_{n\ge 1}
(D^{\wh\cV}_\mu v_\nu^n-D^{\wh\cV}_\nu v_\mu^n)\psi_{2n-1}\nn\\
&&+\sum_{n\ge 1}([\wh\cA_\mu,a_\nu^n]-[\wh\cA_\nu,a_\mu^n])
\psi_0\psi_{2n}\nn\\
&&+\sum_{n,m\ge 1}[v_\mu^n,v_\nu^m]\psi_{2n-1}\psi_{2m-1}
+\sum_{n,m\ge 1}[a_\mu^n,a_\nu^m]\psi_{2n}\psi_{2m}\ ,
\end{eqnarray}
\begin{eqnarray}
F_{\mu\nu}\Big|_{\rm odd}&=&
(D^{\wh\cV}_\mu \wh\cA_\nu-D^{\wh\cV}_\nu \wh\cA_\mu)\psi_0+
\sum_{n\ge 1}(D^{\wh\cV}_\mu a_\nu^n-D^{\wh\cV}_\nu a_\mu^n)
\psi_{2n}\nn\\
&&+\sum_{n\ge 1}([\wh\cA_\mu,v_\nu^n]
-[\wh\cA_\nu,v_\mu^n])
\psi_0\psi_{2n-1}\nn\\
&&+\sum_{m,n\ge 1}([v^n_\mu,a_\nu^m]-[v^n_\nu,a_\mu^m])
\psi_{2m}\psi_{2n-1}\ ,
\end{eqnarray}
where $F_{\mu\nu}\Big|_{\rm even}$ and 
$F_{\mu\nu}\Big|_{\rm odd}$ denote the parts
that are even and odd under $z\ra -z$, respectively,
and we have defined
\begin{eqnarray}
F_{\mu\nu}^{\wh\cV}\equiv
\del_\mu \wh\cV_\nu-\del_\nu \wh\cV_\mu+[\wh\cV_\mu,\wh\cV_\nu]\ ,
~~~D^{\wh\cV}_\mu * \equiv
\del_\mu+[\wh\cV_\mu,*]\ .
\end{eqnarray}
The following are useful relations here:
\begin{eqnarray}
F_{\mu\nu}^{\wh\cV}+[\wh\cA_\mu,\wh\cA_\nu]\psi_0^2
&=&\half(\xi_+ F_{\mu\nu}^{A_L}\xi_+^{-1}
+\xi_- F_{\mu\nu}^{A_R}\xi_-^{-1})
-[\wh\cA_\mu,\wh\cA_\nu](1-\psi_0^2)\ ,\\
D^{\wh\cV}_\mu \wh\cA_\nu-D^{\wh\cV}_\nu \wh\cA_\nu
&=&\half(\xi_+ F_{\mu\nu}^{A_L}\xi_+^{-1}
-\xi_- F_{\mu\nu}^{A_R}\xi_-^{-1})\ .
\end{eqnarray}

The action (\ref{DBI}) is calculated as follows.
The second term in (\ref{DBI}) is
\begin{eqnarray}
\kappa\int\! dz\,K \tr F_{z\mu}^2
&=&
\tr \left[\,
\frac{4}{\pi}\kappa\,\wh\cA_\mu^2+m_{v^n}^2 (v^n_\mu)^2+
m_{a^n}^2 (a^n_\mu)^2
\right]\nn\\
&=&\tr \Bigg[\,
\frac{f_\pi^2}{4} (U^{-1}\del_\mu U)^2+
m_{v^n}^2 (v^n_\mu)^2+m_{a^n}^2 (a^n_\mu)^2\nn\\
&&+\frac{f_\pi^2}{4}\big(
A_{L\mu}^2+A_{R\mu}^2-2U^{-1}A_{L\mu}UA_R^\mu\nn\\
&&~~~~-2A_L^\mu U\del_\mu U^{-1}-2A_R^\mu U^{-1}\del_\mu U
\big)
\Bigg]\ ,
\end{eqnarray}
where we have used the relation (\ref{fpi}).
The first term in (\ref{DBI}) is more complicated.
We segregate it into terms of equal orders in
the vector meson fields $(v^n_\mu,a^n_\mu)$ as
\begin{eqnarray}
\kappa\int\! dz\,
\half K^{-1/3} \tr F_{\mu\nu}^2
&=&\kappa\int dz\, \half K^{-1/3}\tr\left[
 F_{\mu\nu}\Big|_{\rm even}^2+ F_{\mu\nu}\Big|_{\rm odd}^2\right]\nn\\
&\equiv&\cL_{(a,v)^0}+\cL_{(a,v)^1}
+\cL_{(a,v)^2}+\cL_{(a,v)^3}+\cL_{(a,v)^4}\ ,
\end{eqnarray}
where $\cL_{(a,v)^m}$ denotes the terms with $m$ vector meson
fields $v^n_\mu$ and $a^n_\mu$.
The order zero terms are
\begin{eqnarray}
\cL_{(a,v)^0}&=&\kappa\int\! dz\, \half K^{-1/3}\tr\Bigg[
\left(\half(\xi_+ F_{\mu\nu}^{A_L}\xi_+^{-1}
+\xi_- F_{\mu\nu}^{A_R}\xi_-^{-1})
-[\wh\cA_\mu,\wh\cA_\nu](1-\psi_0^2)\right)^2\nn\\
&&~~+\left(
\half(\xi_+ F_{\mu\nu}^{A_L}\xi_+^{-1}
-\xi_- F_{\mu\nu}^{A_R}\xi_-^{-1})
\right)^2\psi_0^2
\Bigg]\nn\\
&=&\tr\Bigg[
\frac{1}{2e^2}\left(
(F_{\mu\nu}^{A_L})^2+(F_{\mu\nu}^{A_R})^2\right)
+\frac{b_{\cV\pi\pi}}{4} U^{-1}F_{\mu\nu}^{A_L}UF^{A_R\,\mu\nu}
\nn\\
&&~~~-\frac{b_{\cV\pi\pi}}{2}(\xi_+ F_{\mu\nu}^{A_L}\xi_+^{-1}
+\xi_- F_{\mu\nu}^{A_R}\xi_-^{-1})
[\wh\cA^\mu,\wh\cA^\nu]
+\frac{1}{2 e_S^2} [\wh\cA_\mu,\wh\cA_\nu]^2
\Bigg] \ ,
\end{eqnarray}
where
\begin{eqnarray}
&&e^{-2}\equiv\frac{\kappa}{4}\int\! dz\, K^{-1/3}(1+\psi_0^2) \ ,\nn\\
&&b_{\cV\pi\pi}\equiv\kappa\int\! dz\, K^{-1/3}(1-\psi_0^2)\ ,~~
e_S^{-2}\equiv\kappa\int\! dz\, K^{-1/3}(1-\psi_0^2)^2\ .
\end{eqnarray}
Here, $e^{-2}$ is divergent, and we should cut off the $z$ integral
to make it finite. Hence, the divergent part $\cL_{\rm div}$
consists simply of the kinetic terms of $A_{L\mu}$ and $A_{R\mu}$:
\begin{eqnarray}
\cL_{\rm div}=\frac{1}{2e^2}\tr \left[
(F_{\mu\nu}^{A_L})^2+(F_{\mu\nu}^{A_R})^2\right]\ .
\label{Ldiv}
\end{eqnarray}

The terms linear in $v^n_\mu$ or $a^n_\mu$ are
\begin{eqnarray}
\cL_{(a,v)^1}
&=&
\tr\Bigg[
\left(\half(\xi_+ F^{A_L\,\mu\nu}\xi_+^{-1}
+\xi_- F^{A_R\,\mu\nu}\xi_-^{-1})
\right)\nn\\
&&~~~~~~~\times
\left(
(D_\mu^{\wh\cV}v_\nu^n-D_\nu^{\wh\cV}v_\mu^n)\, a_{\cV v^n}
+([\wh\cA_\mu,a_\nu^n]-[\wh\cA_\nu,a_\mu^n])\, a_{\cA a^n}
\right)\nn\\
&&~~~-
[\wh\cA^\mu,\wh\cA^\nu]
\left(
(D_\mu^{\wh\cV}v_\nu^n-D_\nu^{\wh\cV}v_\mu^n)\, b_{v^n\pi\pi}
+([\wh\cA_\mu,a_\nu^n]-[\wh\cA_\nu,a_\mu^n])\, b_{a^n\pi\pi\pi}
\right)\nn\\
&&~~~+\left(
\half(\xi_+ F^{A_L\,\mu\nu}\xi_+^{-1}
-\xi_- F^{A_R\,\mu\nu}\xi_-^{-1})
\right)\nn\\
&&~~~~~~~\times\left(
(D^{\wh\cV}_\mu a_\nu^n-D^{\wh\cV}_\nu a_\mu^n)\,
a_{\cA a^n}+
([\wh\cA_\mu,v_\nu^n]-[\wh\cA_\nu,v_\mu^n])\,
(a_{\cV v^n}-b_{v^n\pi\pi})\right)
\Bigg]\ ,\nn\\
\end{eqnarray}
where
\begin{eqnarray}
&&a_{\cV v^n}\equiv\kappa\int\! dz\,K^{-1/3} \psi_{2n-1}\ ,~~~
a_{\cA a^n}\equiv\kappa\int\! dz\,K^{-1/3} \psi_0\psi_{2n}\ ,\\
&&
b_{v^n\pi\pi}\equiv
\kappa\int\! dz\,K^{-1/3} \psi_{2n-1}(1-\psi_0^2)\ ,~~~
b_{a^n\pi\pi\pi}\equiv
\kappa\int\! dz\,K^{-1/3}\psi_0\psi_{2n}(1-\psi_0^2)\ .
\end{eqnarray}

The terms quadratic in $(v^n_\mu,a^n_\mu)$ are
\begin{eqnarray}
\cL_{(a,v)^2}
&=&\tr\Bigg[\,
\half(D_\mu^{\wh\cV}v_\nu^n-D_\nu^{\wh\cV}v_\mu^n)^2\nn\\
&&~~~+\half
([\wh\cA_\mu,a_\nu^n]-[\wh\cA_\nu,a_\mu^n])
([\wh\cA^\mu,a^{m\,\nu}]-[\wh\cA^\nu,a^{m\,\mu}])\,
c_{a^n a^m\pi\pi}\nn\\
&&~~~+
(D_\mu^{\wh\cV}v_\nu^n-D_\nu^{\wh\cV}v_\mu^n)
([\wh\cA^\mu,a^{m\,\nu}]-[\wh\cA^\nu,a^{m\,\mu}])\,
c_{v^na^m\pi}\nn\\
&&~~~+
\half\left(\xi_+ F^{A_L\,\mu\nu}\xi_+^{-1}
+\xi_- F^{A_R\,\mu\nu}\xi_-^{-1}\right)
\left([v_\mu^n,v_\nu^n]+[a_\mu^n,a_\nu^n]
\right)\nn\\
&&~~~-
[\wh\cA^\mu,\wh\cA^\nu]
\left([v_\mu^n,v_\nu^m](\delta_{nm}-c_{v^n v^m\pi\pi})
+[a_\mu^n,a_\nu^m](\delta_{nm}-c_{a^n a^m\pi\pi})
\right)\nn\\
&&~~~+\half
(D^{\wh\cV}_\mu a_\nu^n-D^{\wh\cV}_\nu a_\mu^n)^2\nn\\
&&~~~+\half
([\wh\cA_\mu,v_\nu^n]-[\wh\cA_\nu,v_\mu^n])
([\wh\cA^\mu,v^{m\,\nu}]-[\wh\cA^\nu,v^{m\,\mu}])\,
c_{v^nv^m\pi\pi}\nn\\
&&~~~
+(D^{\wh\cV}_\mu a_\nu^m-D^{\wh\cV}_\nu a_\mu^m)
([\wh\cA^\mu,v^{n\,\nu}]-[\wh\cA^\nu,v^{n\,\mu}])\,
c_{v^n a^m\pi}\nn\\
&&~~~+
\half\left(\xi_+ F^{A_L\,\mu\nu}\xi_+^{-1}
-\xi_- F^{A_R\,\mu\nu}\xi_-^{-1}\right)
([v^n_\mu,a_\nu^m]-[v^n_\nu,a_\mu^m])\,
c_{v^n a^m\pi}
\Bigg]\ ,
\end{eqnarray}
where
\begin{eqnarray}
&&c_{v^n a^m\pi}\equiv
\kappa\int\! dz\,K^{-1/3}\psi_0\psi_{2n-1}\psi_{2m}\ ,\nn\\
&&c_{a^n a^m\pi\pi}\equiv
\kappa\int\! dz\,K^{-1/3}\psi_0^2\psi_{2n}\psi_{2m}\ ,\nn\\
&&c_{v^n v^m\pi\pi}\equiv
\kappa\int\! dz\,K^{-1/3}\psi_0^2\psi_{2n-1}\psi_{2m-1}\ .
\end{eqnarray}

Similarly, $\cL_{(a,v)^3}$ and $\cL_{(a,v)^4}$ are 
\begin{eqnarray}
\cL_{(a,v)^3}
&=&
\tr\Bigg[\,
(D_{\mu}^{\wh\cV}v^n_\nu-D_{\nu}^{\wh\cV}v^n_\mu)
\left(
[v^{p\,\mu},v^{q\,\nu}]\,g_{v^nv^pv^q}
+[a^{p\,\mu},a^{q\,\nu}]\,g_{v^na^pa^q}
\right)\nn\\
&&~~~+
\left(
[\wh\cA_\mu,a^n_\nu]-[\wh\cA_\nu,a^n_\mu]
\right)\left(
[v^{p\,\mu},v^{q\,\nu}]\,g_{a^nv^pv^q}
+[a^{p\,\mu},a^{q\,\nu}]\,g_{a^na^pa^q}
\right)\nn\\
&&~~~+\left(
(D_{\mu}^{\wh\cV}a^n_\nu-D_{\nu}^{\wh\cV}a^n_\mu)\,g_{v^pa^na^q}
+([\wh\cA_\mu,v^n_\nu]-[\wh\cA_\nu,v^n_\mu])\,g_{a^qv^pv^n}
\right)\nn\\
&&~~~~~~\times([v^{p\,\mu},a^{q\,\nu}]-[v^{p\,\nu},a^{q\,\mu}])\Bigg]\ ,
\end{eqnarray}
and
\begin{eqnarray}
\cL_{(a,v)^4}
&=&
\tr\Bigg[\,
\half[v^m_\mu,v^n_\nu][v^{p\,\mu},v^{q\,\nu}]\,g_{v^mv^nv^pv^q}
+\half[a^m_\mu,a^n_\nu][a^{p\,\mu},a^{q\,\nu}]\,g_{a^ma^na^pa^q}
\nn\\
&&~~~
+\left([v^m_\mu,v^n_\nu][a^{p\,\mu},a^{q\,\nu}]+
[v^m_\mu,a^p_\nu][v^{n\,\mu},a^{q\,\nu}]-
[v^m_\mu,a^p_\nu][v^{n\,\nu},a^{q\,\mu}]
\right)
\,g_{v^mv^na^pa^q}
\Bigg]\ ,\nn\\
\end{eqnarray}
where
\begin{eqnarray}
g_{v^nv^pv^q}&\equiv&
\kappa\int\! dz\, K^{-1/3}\psi_{2n-1}\psi_{2p-1}\psi_{2q-1}\ ,\nn\\
g_{v^na^pa^q}&\equiv&
\kappa\int\! dz\, K^{-1/3}\psi_{2n-1}\psi_{2p}\psi_{2q}\ ,\nn\\
g_{a^nv^pv^q}&\equiv&
\kappa\int\! dz\, K^{-1/3}\psi_0\psi_{2n}\psi_{2p-1}\psi_{2q-1}
\ ,\nn\\
g_{a^na^pa^q}&\equiv&
\kappa\int\! dz\, K^{-1/3}\psi_0\psi_{2n}\psi_{2p}\psi_{2q}
\ ,\nn\\
g_{v^mv^nv^pv^q}&\equiv&
\kappa\int\! dz\, K^{-1/3}\psi_{2m-1}\psi_{2n-1}\psi_{2p-1}\psi_{2q-1}
\ ,\nn\\
g_{a^ma^na^pa^q}&\equiv&
\kappa\int\! dz\, K^{-1/3}\psi_{2m}\psi_{2n}\psi_{2p}\psi_{2q}
\ ,\nn\\
g_{v^mv^na^pa^q}&\equiv&
\kappa\int\! dz\, K^{-1/3}\psi_{2m-1}\psi_{2n-1}\psi_{2p}\psi_{2q}\ .
\label{def:g}
\end{eqnarray}

\section{Calculation of $\int_5\omega_5(A)$}
\label{WZW:mani}

Here we outline the derivation of (\ref{omega}).
Because we are working in the $A_z=0$ gauge, we have
\begin{eqnarray}
\omega_5(A)=\tr\left(
A dA dA+\frac{3}{2} A^{3}dA\right)\ .
\end{eqnarray}
We then find
\begin{eqnarray}
\int_5 \tr A dA dA
&=&\int_5\tr\left((v+a) d(v+a) d(v+a)\right)\nn\\
&=&\int_5\tr\left(
v\,dv\,da+v\,da\,dv+a\,dv\,dv+a\,da\,da
\right)\nn\\
&=&\int_5\tr\left(-d(v\,dv\,a+v\,a\,dv)
+3\,a\,dv\,dv+a\,da\,da
\right)\nn\\
&=&\half\int_4
\tr((A_+A_--A_-A_+)\,d(A_++A_-))
+\int_5\left(3\,a\,dv\,dv+a\,da\,da\right)\ ,
\label{AdAdA}
\end{eqnarray}
\begin{eqnarray}
\int_5\tr A^{3}dA
&=&\int_5 \tr (v+a)^3 d(v+a)\nn\\
&=&\int_5\tr\left( (v^3+v\,a^2+a\,v\,a+a^2 v)\,da
+(a\,v^2+v\,a\,v+v^2 a+a^3)\,dv\right)\nn\\
&=&\int_5 \tr\left(
2(v^2 a+a\,v^2+a^3)\,dv+2\,a\,v\,a\,da
-d(v^3 a +v\,a^3+a\,v\,a^2+a^2 v\,a)\right)\nn\\
&=&2\int_5 \tr\left(
(v^2 a+a\,v^2+a^3)\,dv+a\,v\,a\,da\right)\nn\\
&&~~~~-\frac{1}{8}
\int_4\tr\left((A_++A_-)^3(A_+-A_-)+(A_++A_-)(A_+-A_-)^3
\right) \ .
\label{AAAdA}
\end{eqnarray}
The first line on the right-hand side of (\ref{AAAdA}) contains
some terms that do not include vector mesons:
\begin{eqnarray}
2\int_5\tr(a\,v\,a\,da)
&=&\frac{1}{8}
\int_5\tr\left((A_+-A_-)\psi_0(A_++A_-)(A_+-A_-)
\psi_0d\left((A_+-A_-)\psi_0\right)\right)\nn\\
&&~~~~~+2\int_5\Big[\tr(a\,v\,a\,da)\Big]_{\mbox{\footnotesize non-zero}}
\nn\\
&=&
\frac{1}{12}\int_4\tr\left((A_++A_-)(A_+-A_-)^3\right)
+2\int_5\Big[\tr(a\,v\,a\,da)\Big]_{\mbox{\footnotesize non-zero}}\ .
\label{avada}
\end{eqnarray}
As above, $[\,\cdots\,]_{\mbox{\footnotesize non-zero}}$
extracts the terms that include at least one vector meson.
Then using (\ref{avada}), (\ref{AAAdA}) can be rewritten as
\begin{eqnarray}
\int_5 \tr A^{3} dA
&=&2\int_5 \tr\left(
(v^2 a+a\,v^2+a^3)\,dv+\Big[a\,v\,a\,da\Big]_{\rm non-zero}\right)\nn\\
&&~~~+\frac{1}{3}
\int_4\tr\left[\half A_+A_-A_+A_-+(A_+^3A_--A_-^3A_+)
\right]\ .
\end{eqnarray}
Finally, combining (\ref{AdAdA}) and (\ref{AAAdA}),
we obtain (\ref{omega}).

\end{document}